\documentclass[usenatbib]{mnras}
\usepackage[T1]{fontenc}

\usepackage[T1]{fontenc}
\usepackage{ae,aecompl}
\usepackage{graphicx}
\usepackage{hyperref}
\usepackage{amsmath}
\usepackage{cleveref}
\usepackage{amssymb}
\usepackage{mathtools}
\usepackage{bm}
\usepackage{tensor}
\usepackage{braket}
\usepackage{mhchem}

\newcommand{\vect}[1]{\mathbf{{#1}}}
\newcommand{\spc}{\quad \quad \quad}
\newcommand{\paral}{\, \mathbin{\!/\mkern-5mu/\!} \, }

\title[Thermodynamics of relativistic multifluids]
{Thermodynamics of uncharged relativistic multifluids}

\author[L. Gavassino \& M.Antonelli]{
L.~Gavassino, 
M.~Antonelli
\\ \\
Nicolaus Copernicus Astronomical Center of the Polish Academy of Sciences, Bartycka 18, 00-716 Warszawa, Poland
}

\begin{document}



\maketitle

\label{firstpage}

\begin{abstract}

The internal layers of neutron stars are expected to contain several superfluid components that can significantly  affect their dynamics. The description of such objects should rely on hydrodynamic models in which it is possible to unambiguously assign the value of the thermodynamic variables from microscopic calculations of the properties of matter.
In this work we consider the phenomenological approach to multifluids modelling championed by Carter and,  studying the relaxation of the system towards equilibrium, we assign a precise thermodynamic interpretation to its variables.
We show that in thermodynamic equilibrium the equation of state contains less state variables than those needed in the phenomenological model, implying the existence of a gauge freedom of the theory that can be  used to simplify the hydrodynamic formulation in the non-dissipative limit. 
Once this is understood, it becomes easy to translate the different multifluid formalisms that have been proposed in the literature  into Carter's form.
Finally, we show that the usual concepts of affinity and reaction coordinates, as they are introduced in chemistry, are not affected by the presence of superfluid currents. In an effort to make the concepts clear, the formalism  is developed  step-by-step from first principles, providing model examples and several 
applications of practical relevance for the study of superfluid neutron star interiors.

\vspace{10mm}
\end{abstract}



\section{Introduction}
\label{sec:intro}

%
%
%
%
%
Superfluidity in neutron star interiors opens the door to a wide range of exotic scenarios where  metastable states in which two or more persistent currents can flow with respect to each other can be sustained for an extremely long time \citep{sauls1989superfluidity,haskellsedrakian2017,chamel_super}. 
Therefore, we are forced to extend in a relativistic context the notion of fluid to include non-divisible substances which are crossed by more than one independent currents, namely a multifluid. The dynamics of  multifluids is intrinsically different from that of ordinary mixtures, since in the former case the fluid components may have a different four-velocity also in the non dissipative limit \citep{andersson2007review}. 
Apart from a  richer dynamics, this fundamental difference introduces new properties also at the thermodynamic level since equilibrium states supporting persistent currents are possible, and observed, in terrestrial superfluid systems \citep{reppy_64}.

The Lagrangian approach to multifluid hydrodynamics championed by B. Carter and coworkers  provides a powerful covariant formalism to construct phenomenological multifluid models \citep{noto_rel}. 
However, our ability to reproduce the dynamics of a realistic system rests ultimately on the equation of state, which is usually a product of some microscopic calculations that are carried out in a particular frame. 
Since multifluid models in General Relativity must be considered for an accurate modelling of superlfuid neutron star interiors, the problem of how to embed the information given by a microphysical equation of state within a covariant phenomenological model is of primary importance: the wide range of applications spans from equilibrium models of stellar structure \citep[e.g. ][]{andersson_comer2000,sourie2016PhRvD}, to the study of stellar oscillations \citep[e.g. ][]{and_com_lang_2002,GusakovAndersson2006}, pulsar glitches in General Relativity \citep[e.g. ][]{sourie_glitch2017,antonelli+2018} and cooling models \citep{yako_cooling_rev}.
%

Leaving aside the major theoretical concern about how to include in a covariant formalism dissipative effects \citep[see e.g. ][ for an approach to the issue within the multi-fluid paradigm]{lopez2011} and elastic components  \citep{carter1972foundations,ander_elasticity_2019} needed to describe the crust of neutron stars \citep{chamel_review_crust}, also the  non-dissipative limit of the theory may challenge the current understanding when applied to systems containing several species. In fact, we find that a  formalism based on a master function allows for a gauge freedom when several non-superfluid species are present: the sophistication of the action principle, seen as the machinery which encodes the dynamics of the multifluid in the adiabatic limit, gives rise to a separation between the physical quantities and some redundancies which are required at the mathematical level. Our aim is to provide a physically motivated thermodinamic language that is gauge invariant.
\\
\\
The paper is organized as follows.

In section 2 we recall the variational approach of \cite{Comer_1993}, see also \cite{andersson2007review}, to set the notation and introduce a general phenomenological model for the dynamics of a perfect (i.e. non-dissipative) multifluid.

Section 3 is devoted to deriving all the equilibrium properties of a multifluid in contact with a heat bath, using only the expression of the energy-momentum tensor derived in section 2, the second law of thermodynamics and the irrotationality of the superfluid momentum.

In section 4 we use  the tetrad formalism to recast the equations of non-dissipative multifluid hydrodynamics in a way that only variables with well defined thermodynamic interpretation appear; this reveals the connection between the Lagrangian approach and the one developed by \cite{Son2001} and later generalized by  \cite{GusakovAndersson2006}.

In section 5 we give an example of how it is possible to employ our thermodynamic approach to model dissipative phenomena, studying the problem of the relaxation to chemical equilibrium in a generic multifluid and showing that the two formulations of \cite{langlois98} and \cite{Gusakov2007} are equivalent. 
We analize how the so-called ``rocket-effect''  changes under a gauge transformation and use this result to reduce its form to the one proposed by  \cite{ColucciRocket2013} in the low temperature limit.

The appendices are devoted to discussing of some issues of practical relevance for neutron star modelling. In particular, in appendix C we solve a dictionary problem between two different formulations for the hydrodynamics in the outer core of a neutron star developed by \cite{pca2002} and \cite{KP2017}.
\\
\\
Throughout the paper we adopt the spacetime signature $ ( - , +, + , + ) $, choose units with the speed of light $c=1$ and Newton's constant $G=1$, use greek letters $\nu$, $\rho$, $\sigma$... for coordinate tensor indices.

\section{Relativistic multifluid  model}
\label{sec:Gftva}

In this section, we briefly review the variational principle for non-dissipative multifluid hydrodynamics introduced in \cite{Comer_1993} and \cite{comer_langlois1994}. This allows us to set up notation and to remark some subtleties (like the distinction between the Lagrangian and the master function for the system) which will be important in the following sections. 

\subsection{The variational approach for non-dissipative models}
\label{Gftva}

The fundamental ingredients in a multifluid theory are the currents, one for each of the chemical species comprising the system. Those currents are vector fields whose norm represents the local rest-frame density of the relative chemical species\footnote{
	In the present context the term ``chemical species'' denotes a subset of the particles comprising the substance, which are grouped according to a non-specified criterion. From a general point of view, it is not necessary that all the particles present in the system contribute to the currents (e.g. the electrons in \cite{langlois98}), nor a multiple counting of the same particle is forbidden (e.g. the density of ions in \cite{carter_macro_2006}). However, the full set of currents (entropy included) should give  without redundancies the complete information about the local state of matter. This is related to the ``chemical gauge'' problem discussed in \cite{carter_macro_2006}, see appendix \ref{Chemical gauge in an arbitrary perfect multifluid}.
}
and whose direction is collinear to its average four-velocity. 
An additional ``thermal current'' $s^\nu$ carries information about the local frame, identified by the four-velocity $u^\nu$ say, in which the excitations present in the system thermalize (i.e. have average zero velocity) and about the entropy of the system. Hence, it is always possible to write the thermal current as
\begin{equation}
s^\nu = s u^\nu \, ,
\label{u_T}
\end{equation}
where the scalar $s=\sqrt{-s_\mu s^\mu}$ is the entropy per unit volume measured in the local aforementioned frame.

For reasons which will become clear in the following section we assume that the remaining independent currents can be divided into two subsets. We suppose that a number $l$ of them are forced to comove with the entropy, so we call them ``s-locked'' and count them with the labels $A,B$ running from 0 to $l$ (0 identifies the thermal current).  According to this notation, the s-locked currents can be written as
\begin{equation}\label{s-locking}
n_A^\nu = n_A u^\nu \, .
\end{equation}
Since there is no net entropy flow in the frame identified by $u^\mu$, no heat transport is measured by an observer comoving with the s-locked components.
 
We assume the remaining $k$ currents to be completely independent; the labels $i,h$ (running from 1 to $k$) are used to count them. 
Since they can flow in an arbitrary direction we call them ``free'' and in general we have 
\begin{equation}
n_i^\nu = n_i u_i^\nu \, ,
\end{equation}
where $u_i^\nu$ can be different from $u^\nu$. No Einstein summation convention will be used for species labels $A,B, i, h$.

Since we have a total number of $1+l+k$ currents, there are  $4(1+k+l)$ variables that may be used to construct a Lagrangian for the system. However, only $l+4(1+k)$ of such variables are independent, namely $s^\nu$, $n_i^\nu$ and the scalars $n_A$. Despite the fact that these are all the degrees of freedom required to define completely the local state of the multifluid at a point, the fields upon which the action will depend do not need to be so numerous. 

The equations of motion in a spacetime region $\mathcal{M}$ have to be found by imposing the stationarity of the total action $I$ with respect to a variation of the fundamental fields of the theory \citep{noto_rel}. As usual, $I$ is the sum of a matter term $I_m$ and of the Einstein-Hilbert action
\begin{equation}
I_{EH} = \int_\mathcal{M} \dfrac{R}{16 \pi} \sqrt{-g} \, d_4 x \, ,
\end{equation}
where $R$ is the scalar curvature. A boundary term, unessential in the present discussion, is also necessary to give a well posed action principle \citep[see e.g. ][]{andersson2007review}. In this way the Euler-Lagrange equations associated to a variation of the components of the metric are
\begin{equation}
G_{\nu \rho} = 8 \pi T_{\nu \rho} \, ,
\end{equation} 
where $G_{\nu \rho}$ is the ususal Einstein tensor and
\begin{equation}
T_{ \nu\rho} \, = \, -\dfrac{2}{\sqrt{-g}} \dfrac{\delta I_m}{\delta g^{ \nu \rho}} \, .
\end{equation}
%
Since we are specializing our study to a non-dissipative and non-transfusive model, we follow the variational procedure proposed by \cite{Comer_1993} to ensure the conservation of the number density currents. In particular, the conservation laws
\begin{equation}
\label{continuityyyy}
\nabla_\nu n_x^\nu=0 \, ,
\end{equation}
where the label $x$ runs over all the currents (entropy included), should be automatically satisfied both on shell (i.e. on the solutions of the equations of motion) and off shell (i.e. on a generic spacetime-matter configuration which belongs to the domain of the action). 
This is done by realizing that the number of particles of type $x$ contained in an infinitesimal volume element transported by the four-velocity field $u_x^\nu$ is conserved: all the information about the current is contained, once the initial condition is assigned, into three scalar fields giving the comoving (Lagrangian) coordinates of the volume element occupying the point in which they are evaluated \citep{Comer_1993, andersson2007review}. 
Therefore, we impose that the action depends only on the 3 scalar fields $X_s^\alpha$, for $\alpha=1,2,3$, describing the worldline congruence associated to $u^\nu$ and 3 more scalar fields $X_i^\alpha$ for every free component. In the end, apart from the metric, the total action $I$ is a functional of $3(1+k)$ independent fields:
\begin{equation}
I \, = \,I_{EH}[\, g \,] + \, I_m[ \, g \, , \, X_s^\alpha \, , \, X^\alpha_{i} \, ] \, .
\label{anchilosauro}
\end{equation}
In this way the s-locked components comove with the entropy both on shell and off shell. This will be particularly important in section \ref{Mathematical explanation of the redundancy}.

\subsection{The master function and the equations of motion}
\label{variationalprinciplesubsection}

%
In the previous section we introduced the basis for the most general multifluid theory where all the components are uncharged and there are no dissipative phenomena. The Lagrangian $\mathcal{L}$ for the multifluid is introduced by imposing that the matter action has the local  form
\begin{equation}
\label{matteraction}
I_m = \int_\mathcal{M} \mathcal{L} \sqrt{-g} \, d_4 x \, .
\end{equation}
A general multifluid theory can be obtained by taking every species of the theory, no matter if they are independent or not, and constructing the symmetric matrix $n_{xy}^2:= -n_x^\nu n_{y\nu}$. 
We consider only the upper triangle $x \leq y$ comprised of 
\begin{equation}
z=\dfrac{(1+k+l)(2+k+l)}{2}
\end{equation} 
terms and introduce the  \emph{master function}
\begin{equation}
\Lambda \, : \, \mathbb{R}^z \, \longrightarrow \, \mathbb{R} \, ,
\end{equation} 
which is just an auxiliary function (a sort of pre-Langrangian).
At this point, the Lagrangian density is obtained from $\Lambda$ as
\begin{equation}
\label{ilbuglagrangiano}
\mathcal{L}[  g  , X_s^\alpha  , X^\alpha_{i}  ]
= 
\Lambda ( -g_{\rho \nu} \, n_x^\rho[ g  , X_x^\alpha  ] \, n_y^\nu[  g  , X_y^\alpha  ] ) 
\, .
\end{equation}
We make this distinction between $\mathcal{L}$ and $\Lambda$, which is usually not found in literature, to remark that what matters for the dynamics is the value of $\mathcal{L}$ on shell and off shell. On the other hand,  
$\Lambda$ is an intermediate function that can receive arguments which are incompatible with the constraint of comoving s-locked components. 

We assume $\Lambda$ to be differentiable, which means that we are ignoring the possibility of first order phase transitions. We define the \textit{bulk} coefficients 
\begin{equation}
\label{bcoeff}
\mathcal{B}^x := -2 \dfrac{\partial \Lambda}{\partial n_{xx}^2} \, ,
\end{equation} 
where, following \cite{carter92}, we adopt the notation $\mathcal{B}^s=\mathcal{C}$ for the one associated to the entropy. Furthermore, the so-called \textit{anomaly} coefficients can be casted into a symmetric matrix 
$\mathcal{A}^{xy}$, whose upper triangle is defined as
\begin{equation}
\label{anomaly1}
\mathcal{A}^{xy}:= -\dfrac{\partial \Lambda}{\partial n^2_{xy}} 
\qquad for  \qquad
x < y \,.
\end{equation}
%
%
The coefficients $\mathcal{A}^{xy}$ are responsible for the \emph{entrainment} effect when non-comoving species are present in the system \citep[][]{khala57,andreevbashkin1976}. 
The partial derivatives are here interpreted in a strict sense: they are derivatives of the master function $\Lambda$ with respect to its argument $n_{xy}^2$. 
However, if $\Lambda$  is seen as a function of all the components of the $1+l+k$ currents \emph{independently} (i.e. regardless of the fact that the s-locked currents are forced to comove with the entropy), then it can be seen as a function of $4(1+l+k)$ variables through the formula 
\begin{equation}
\Lambda (n_{x}^\nu) := \Lambda(-g_{\rho \nu} n_x^\rho n_y^\nu) \, .
\end{equation}
Thanks to this operation, the canonically conjugate momenta are defined as
\begin{equation}
\label{noivinciamosempre}
\mu_{\nu}^x
:= 
\dfrac{\partial \Lambda}{\partial n_x^\nu} 
= 
\mathcal{B}^x n_{x\nu} + \sum_{y \neq x} \mathcal{A}^{xy}n_{y\nu} \, , 
\end{equation}
where the sum runs over both the upper and lower triangles.
Again, we stick to Carter's notation and the momentum $\mu_{\nu}^s$ related to the entropy current is denoted by $\Theta_\nu$.

Following a derivation that is completely analogous to the one delineated e.g. in \cite{andersson2007review}, it is possible to prove that the energy-momentum tensor is given by
\begin{equation}
\label{energiamomentotempfinita}
T\indices{^\nu _\rho} \, = \, \Psi \delta\indices{^\nu _\rho} + \sum_x n^\nu_x \mu_{\rho}^x \, ,
\end{equation}
where the generalized pressure\footnote{
The generalized pressure can be defined by considering a stationary flow trough a channel: it is the pressure exerted by the fluid on the channel walls. To see this it is sufficient to take an orthonormal basis $e_a$ such that $e_1$ is normal to the wall: since $n_x^1=0$ (otherwise the $x$ component would flow through the wall), we have $T_{11}=\Psi$.} $\Psi$ is related to the master function via
\begin{equation}\label{pressionciona}
\Psi = \Lambda - \sum_x n^\rho_x \mu_{\rho}^x \, .
\end{equation}
Finally, the hydrodynamic equations are
\begin{align}
\begin{split}
& n_{i}^\rho \partial_{[\rho}  \, \mu_{ \nu]}^i = 0  \qquad for \qquad i=1,...,k
\\
&   \sum_{A=0}^l  n_{A}^\rho \partial_{[\rho}  \, \mu_{ \nu]}^A = 0 
\end{split}
\label{equationsofmotioningeneral}
\end{align} 
Once $\nabla_\nu n_x^\nu=0$ is satisfied, the above system of equations implies that $\nabla_\rho T^{\rho \nu}=0$, an information which is already contained in the Einstein's equations: the Euler-Lagrange equations associated to the worldline congruence of the entropy is  redundant and could be replaced by the conservation of the total energy-momentum tensor.

\section{Equilibrium properties of homogeneous systems}
\label{Equilibrium properties of homogeneous systems}

%

The multifluid model of section \ref{sec:Gftva} is completely phenomenological. 
%
In this section we show how to derive a thermodynamic description which is consistent with this model.
The following discussion defines, starting from first principles, the connection of the phenomenological theory with the microphysics of the system and provides a new thermodynamic interpretation of many results which in other treatments required the inclusion of dissipation in order to be concretely proved.

\subsection{Equilibrium with a heat bath}
\label{EwHb}

In thermodynamics it is conventional to consider a homogeneous macroscopic portion of the given substance, small enough to neglect gravity. To implement invariance under translations of the equilibrium state, while working with finite extensive quantities, the fluid is enclosed into a cubic volume $V=L^3$ with periodic boundary conditions. This box is a test spacetime, whose topology is $ \mathbb{R} \times \mathbb{T}^3 $, equipped with a chart $(t,x^1,x^2,x^3)$ and the Minkowskian metric $\eta=\text{diag}(-1,1,1,1)$.
%
%
The periodic boundary conditions read $\{ x^j =0 \} \equiv \{ x^j =L \}$ for $j=1,2,3$.
It is useful to assume that the box is in contact with an ideal heat bath of constant temperature $\Theta_H$, at rest in the global frame defined by the aforementioned chart, i.e. an effectively infinite mass-energy reservoir characterized by the equation of state
\begin{equation}
E_H (S_H) = E_{H0} + \Theta_H S_H \, ,
\end{equation}
where $E_H$ is the energy measured by an observer $\mathcal{O}$ such that $u_{\mathcal{O}}=\partial_t$. 
Imagine now to put a sub-system $M$ inside the box, with an arbitrary initial condition: after an equilibration process involving a complex dissipative evolution partially driven by the interaction with the bath, $M$ will reach a state of thermodynamic equilibrium. In particular, the second principle of thermodynamics, together with the conservation of the total energy of $M$ plus the bath, implies that $M$ will reach the state which minimizes the quantity
\begin{equation}\label{dacapoancoraallinizio}
F \, = \, E_M - \Theta_H S_M \, ,
\end{equation}
where $E_M$ and $S_M$ are the energy and the entropy of the subsystem as mesured in the frame of  $\mathcal{O}$. Assuming that we can associate to the substance constituting the system an energy-momentum tensor and that the final equilibrium state is homogeneous, the energy of $M$ can be identified as 
\begin{equation}
\label{belloo}
E_M = V T^{00},
\end{equation}
where the superscript $0$ refers to $t$. Similarly, the total entropy $S_M$ is 
\begin{equation}
\label{bello}
S_M = Vs^0 \, ,
\end{equation}
where, according to \eqref{u_T}, $s^0$ is the density of entropy measured by $\mathcal{O}$.

Let us now specialize the analysis to the multifluid case. As a starting hypothesis, the system $M$ is prepared in an arbitrary initial condition, namely all the currents can be considered to be free. Combining equations \eqref{energiamomentotempfinita} and \eqref{pressionciona} we obtain
\begin{equation}
T^{00} = -\Lambda + \sum_x n_x^j \mu^x_j ,
\end{equation}
where $j$ runs over the spatial indices and obeys the Einstein summation convention. Plugging these results inside \eqref{dacapoancoraallinizio} we obtain that the equilibrium state minimizes the function
\begin{equation}\label{energialiberallinizio}
\mathcal{F} = F/V  = -\Lambda + \sum_x n_x^j \mu^x_j -\Theta_H s^0.
\end{equation}
Explicitly, this minimization condition reads (notice the raised index of the zero-th components of the momenta)
\begin{equation}
\label{equilibration}
\delta \mathcal{F} = \sum_x (\mu^{x0}\delta n_x^0 + n_x^j \delta \mu_{j}^x) - \Theta_H \delta s^0 =0
\end{equation}
for any dynamically allowed infinitesimal variation. 

Following \cite{Callen_book}, let us introduce the manifold of homogeneous macrostates $\mathcal{Z}$, namely the set of all the states that can be realized in the box assuming all the currents (entropy included) to be free. A point in  $\mathcal{Z}$, i.e. a generic macrostate, is identified by specifying the values of the variables of state $(n_x^\nu)_{x \in C} $, where $C$ is the set of all the currents. If $\Lambda$ is convex in the variables $n_x^j$, the map
\begin{equation}
\label{biezione}
(n_x^\nu)_{x \in C}  \quad \longmapsto \quad (n_x^0,\mu^x_j)_{x \in C} 
\end{equation}
is a bijection. Therefore,  $(n_x^0,\mu^x_j)_{x \in C} $ is an alternative choice of coordinates on  $\mathcal{Z}$.

In the absence of quantities that are conserved during the equilibration process, the variations $\delta n_x^0$ and $\delta \mu^x_j$ in \eqref{equilibration} are unconstrained and independent with respect to each other, leading to the conditions $\mu^{x0}=0$ for $ x \neq s$. This would give the empty space as a unique solution. However, the presence of constants of motion imposes some constraints on the macrostates that the system can explore and, therefore, on the allowed variations. Only by taking into account these constraints it is possible to have non-trivial equilibria: this is explored in the following two sections. 

\subsection{Equilibration in the non-transfusive limit}
\label{Constraining the variations in the non-transfusive limit}

Unconstrained variations in \eqref{equilibration} give trivial equilibria. On the other hand, non-trivial equilibria are obtained when we restrict the allowed variations on a submanifold in  $\mathcal{Z}$.

As a first case, consider the non-transfusive theory, where the number of particles of each chemical species is conserved. In this case, we have to consider the subspace in  $\mathcal{Z}$ defined by the constraints
\begin{equation}\label{consnx}
n_x^0 = \dfrac{N_x}{V}  \spc  x \neq s,
\end{equation}
where $N_x$ is the number of particles of the species $x$ and its value is imposed in the initial conditions in which the system is prepared before equilibration. 

Apart from the particle number conservation, the presence of superfluid components in the multifluid  introduces other constants of motion, whose conservation is directly linked to the possibility to have persistent currents at equilibrium \citep{reppy_64}. 

Let us suppose that a generic component $x$ is superfluid. Following \cite{carter92}, in the absence of vortices we have to impose that
\footnote{
The irrotationality condition  is not contained into the phenomenological model of the previous section, where the quantities $\mu^x_\nu$ have no assigned physical meaning. To enforce internal consistency, this physical identification with the gradient of a phase has to be done by looking at the equations of motion of the phenomenological model: the first equation in \eqref{equationsofmotioningeneral} tells us that an irrotationality requirement can be imposed on the quantities $\mu^i_\nu$ (and not on other combinations  different from \eqref{noivinciamosempre}).
} 
\begin{equation}\label{nonruotanoapagarli}
\partial_{[\rho} \mu_{\nu]}^x =0 .
\end{equation}
This irrotationality condition is interpreted as resulting from the requirement that
\begin{equation}\label{topologiallariscossa}
\mu^x_\nu = \dfrac{k_x}{2\pi} \partial_\nu \phi^x,
\end{equation}
where $\phi^x$ is the phase of a scalar order parameter for the species $x$ under consideration. Here $k_x$ is equal to $2\pi \hbar$ when $x$ is a Bose fluid, while $\pi \hbar$ in the case of a Fermi fluid. The key assumption \eqref{topologiallariscossa} is best justified by its physical consequences, as discussed by \cite{anderson_considerations}: it is indeed the covariant form of the Josephson relation for neutral superfluids \citep{carter_macro_2006,chamel_jltp_2017}. 

Take now  equation \eqref{nonruotanoapagarli} for, say, $\rho=0$ and $\nu=1$ and integrate it along $x^1$: 
\begin{equation}
\dfrac{d}{dt} \int_0^L \mu_1^x(t,x^1,0,0) dx^1 = \mu_0^x(t,L,0,0) - \mu_0^x(t,0,0,0).
\end{equation}
Periodic boundary conditions imply that the right-hand side vanishes, so that the quantity\footnote{
This quantity could be defined by integrating the momentum along any  path that is homotopically equivalent to a loop embracing the torus in the direction $x^1$ exactly once.
}
\begin{equation}\label{windino}
W_{1}^x :=\dfrac{1}{k_x} \int_0^L \mu_1^x(t,x^1,0,0) dx^1
\end{equation}
must be conserved during the evolution of the system, provided that the interaction with the heat bath does not destroy the order parameter anywhere (i.e. there are no zeroes of the order parameter). This can be also shown with a simple topological argument. Considering \eqref{topologiallariscossa},  $W_1^x$  is the number of windings of the phase $\phi^x$ across the torus along the direction 1. Since $\mu_j^x$ are smooth functions on the spacetime, also $W_1^x$ has to vary continuously with time. However, the only continuous functions from $\mathbb{R}$ to $\mathbb{Z}$ are the constant ones, which implies that $W_1$ is conserved during the equilibration process with the heat bath. 

For homogeneous configurations, the conservation of the winding numbers (similarly one can define also $W_2^x$ and $W_3^x$) implies  the further constraints
\begin{equation}\label{quantizzo per sempre}
\mu_j^x = \dfrac{k_x W^x_j}{L}.
\end{equation}
Therefore, equation \eqref{consnx}  tells us that we have to impose $\delta n_x^0=0$ for any $x$, while 
\eqref{quantizzo per sempre} forces us to set $\delta \mu_j^x=0$ when $x$ refers to a superfluid species. 
With these constraints the condition \eqref{equilibration} becomes
\begin{equation}\label{equilibration2}
 s^j \delta \Theta_j + \sum_{x \in \mathcal{N}}  n_x^j \delta \mu_{j}^x +(\Theta^0 -\Theta_H)\delta s^0  =0 ,
\end{equation}
where $\mathcal{N}$ is the set of the normal (i.e. non-superfluid) species. 
Now, the remaining infinitesimal variations appearing in the left-hand side of the above equation are free and independent with respect to each other, implying that $\delta \mathcal{F}$ is zero only if all the prefactors vanish. Let us examine the equations we get one by one. First of all, we have
\begin{equation}\label{freniamolentropia}
s^j=0  \spc  j=1,2,3.
\end{equation}
This equation states that the excitations of the system thermalize in a way to be on average at rest in the frame of the heat bath. This can also be interpreted by saying that, in conditions of thermal equilibrium, the four-velocity $u^\nu$, defined in \eqref{u_T}, identifies the observer in whose reference frame the multifluid can be studied using the canonical ensemble. 

Secondly, for the normal species we have 
\begin{equation}\label{freniamolenormali}
n_x^j =0 \spc j=1,2,3 \spc x \in \mathcal{N}.
\end{equation}
Due to the absence of conserved quantities other than the particle numbers, a normal component initially flowing with respect to the heat bath will be slowed down by an entropic force arising from the interaction with the bath, until the flow stops. This, however, does not happen to the superfluid currents, which can survive at a thermodynamic level as a result of the conservation of the winding numbers\footnote{
The superfluid momentum is modified  by the motion of physical boundaries or of quantized vortices \citep{donnelly_book}. This may be seen as  the physical manifestation  of the conservation of the winding numbers in an annulus. It is also possible to give a more microscopic interpretation, at least for a bosonic superfluid \citep{NP_book}: as a result of the interaction with the bath, particles may be knocked out of the condensate; this process reduces the amplitude of the order parameter associated to condensation, but it will not alter its phase. Hence, changing the momentum of the superfluid would require a transition that involves a macroscopic number of particles from one state to another simultaneously, a very low probability process.
}  $W^x_j$.
Equations \eqref{freniamolentropia} and \eqref{freniamolenormali} combined imply that thermodynamic equilibrium is reached when all the normal currents comove with the entropy (i.e. they are s-locked). Of course this is not necessarily true out of equilibrium, when dissipation occurs. 

The last equation we get from \eqref{equilibration2} is
\begin{equation}
\label{temperiamolatemperatura}
\Theta^0 = \Theta_H.
\end{equation} 
As expected, when the system is in equilibrium with the bath, its temperature must coincide with $\Theta_H$; together with \eqref{freniamolentropia}, this statement assures that the absolute temperature $\Theta$ of a multifluid in thermodynamic equilibrium can be expressed in a covariant fashion as
\begin{equation}\label{temperiamolatemperatura2}
\Theta = -\Theta_\nu u^\nu .
\end{equation}
Clearly, this coincides with the value $\Theta^0$ measured by an observer comoving with the s-locked components.
 
\subsection{Equilibration in the presence of transfusion}\label{chemicandodurissimooo}

In the previous subsection we have computed the equilibrium properties of a multifluid in the non-transfusive limit. Let us see what happens if we relax this assumption and assume that a reaction 
\begin{equation}\label{lareazioncionagenerica}
\sum_{x \neq s} \alpha_x X_x \ce{ <=> } \sum_{x \neq s} \beta_x X_x
\end{equation}
can occur and that no other ways to modify $N_x$ are allowed. Here $X_x$ is the name of the chemical species associated with the current $x$, while $\alpha_x$ and $\beta_x$ are the stoichiometric coefficients. Let us introduce the \emph{reaction coordinate} $N$, also called in chemistry \emph{extent of reaction}, which counts the number of reactions from the left to the right minus the number of reactions from the right to the left, setting the zero on the initial condition. Starting from a state in which the particle numbers are $N_x^{(in)}$, we have
\begin{equation}
\label{conservotutto}
N_x = N_x^{(in)} + (\beta_x - \alpha_x)N.
\end{equation}
This means that the variations $\delta n_x^0=\delta N_x/V$ are not forced to be zero any more, but they are still not completely free. Instead, they assume the form
\begin{equation}
\label{mnnbufti}
\delta n_x^0 = (\beta_x - \alpha_x)\delta n^0, \spc x\neq s,
\end{equation}
where we have defined\footnote{
For now, the index $0$ of $n^0$ is a pure notation convention. In section \ref{Chemical reactions in almost-perfect multifluids} we give an interpretation of $n^0$ as the zero-th component of a virtual current, giving a formal justification for this convention.
} 
$n^0:= N/V$.
As discussed in the previous section , the conservation of the winding numbers $W^x_j$ is not affected by the reactions. Hence, in this case equation \eqref{equilibration} reduces to
\begin{equation}\label{eqeqeqeqeqchimicissimo}
 s^j \delta \Theta_j + \sum_{x \in \mathcal{N}}  n_x^j \delta \mu_{j}^x +(\Theta^0 -\Theta_H)\delta s^0  
+
\sum_{x \neq s} (\beta_x - \alpha_x)\mu^{x0} \delta n^0 =0 ,
\end{equation}
a form which is identical to \eqref{equilibration2} plus an additional term due to the presence of reactions.

The submanifold of the allowed macrostates is fully described by the variables $(n^0,s^0,\Theta_j,\mu_j^x)_{x \in \mathcal{N}}$, whose variations are independent. 
This means that equations \eqref{freniamolentropia}, \eqref{freniamolenormali}, \eqref{temperiamolatemperatura} and \eqref{temperiamolatemperatura2} remain true, but we get one more condition arising from the presence of the additional term  in \eqref{eqeqeqeqeqchimicissimo},
\begin{equation}
\sum_{x \neq s} \alpha_x \mu^{x0} = \sum_{x \neq s}  \beta_x \mu^{x0}.
\end{equation}
Not surprisingly, this is the conventional formula for the chemical equilibrium. However, we obtained more than a well-known result: as \eqref{temperiamolatemperatura} provides a first-principles justification of the notion of absolute temperature in \eqref{temperiamolatemperatura2}, similarly this relation provides a physical interpretation of the notion of chemical potential. In fact, since \eqref{freniamolentropia} still holds, we define the thermodynamic chemical potentials (i.e. the quantities needed in the equation for chemical equilibrium) as
\begin{equation}\label{chemichiamolachimica}
\mu_x^{(T)} := -\mu^x_\nu u^\nu \, .
\end{equation}
In this way,  the chemical equilibrium condition can be written in terms of covariant quantities as
\begin{equation}\label{tantovalagatta...amorire!}
\sum_{x \neq s} \alpha_x \mu_x^{(T)} = \sum_{x \neq s}  \beta_x \mu_x^{(T)}.
\end{equation}
This result is perfectly coherent with \cite{langlois98}, where beta reactions in superfluid neutron stars are taken into account. 

As a final remark, notice that in the above analysis we have treated $N$ as a completely free variable, without considering the fact that $N_x$ must be non-negative in \eqref{conservotutto}. The variable $N$ should move along a segment whose extrema are reached when we run out of a reactant and it can happen that the minimum of the free energy is reached on the border of this segment, for example when one reaction always dominates over the other and does not stop until all its reactants are transformed into products. In this case equation \eqref{tantovalagatta...amorire!} needs not to be satisfied. 


\subsection{Thermodynamic potentials}
\label{thermopetentissimi}

The thermodynamic properties of a system are completely known once an equation of state is given. Since all the theory describing a multifluid is determined once a master function $\Lambda$ is assigned, it is usually stated that 
\begin{equation}
\Lambda = \Lambda (n_{xy}^2)
\end{equation}
represents the equation of state of the multifluid. Even if $\Lambda$ contains all the information needed to describe the system and can be obtained from an underlying relativistic mean field theory (see \cite{glendenning_book} for the multi-component single fluid, \cite{comerRMF} for the two-component multifluid and appendix \ref{CconRMF} for an extension to the three-component case) its thermodynamic interpretation is not obvious. In this subsection we assign a precise thermodynamic meaning to $\Lambda$. In doing this, we show that all the thermodynamic potentials usually employed in single-fluid thermodynamics can be recovered, and extended, to the multifluid case. 

Consider the set $\mathcal{Q}$ of all the possible equilibrium macrostates in which we can prepare the multifluid by setting the temperature of the bath\footnote{
Clearly, $\mathcal{Q}  \subset \mathcal{Z}$. For a given  temperature of the bath, the equilibration process projects a point in $\mathcal{Z}$ on $\mathcal{Q}$.
}. 
The natural full set of variables which we can use to label a point in $\mathcal{Q}$ is
\begin{equation}
(\Theta_H, N_A,N_i, W^i_j)^{A=1,...,l}_{i=1,...,k} \, .
\end{equation}
This makes $\mathcal{Q}$ a manifold of dimension $1+l+4k$. 
Assuming proper convexity properties for $\Lambda$ and given that the volume $V$ is a constant, the map
\begin{equation}
(\Theta_H, N_A,N_i, W^i_j)^{A=1,...,l}_{i=1,...,k} \quad \longmapsto \quad (n_A^0,n_i^0,n_i^j)^{A=0,...,l}_{i=1,...,k}
\end{equation}
is a bijection, meaning that both sets of variables can be used to label a point in $\mathcal{Q}$. In particular, if we now consider $\Lambda$ as a function from $\mathcal{Q}$ to $\mathbb{R}$, its differential has the form
\begin{equation}
d\Lambda = \sum_{A=0}^l \mu^A_0 dn_A + \sum_{i=1}^k \mu^i_0 dn_i^0 + \sum_{i=1}^k \mu^i_j dn_i^j,
\end{equation}
where, since the normal components are at rest in the frame of the bath, $n_A^0 =n_A$  (no confusion should arise between indexes $i$ and $j$ because the sum over the former is always explicit and for the latter we use the Einstein convention). The internal energy density $\mathcal{U}$ of the multifluid in thermal equilibrium with a bath is defined to be the energy density $T^{00}$ (i.e. measured in the reference frame of the bath) seen as a function on $\mathcal{Q}$. Combining \eqref{energiamomentotempfinita}, \eqref{pressionciona}, \eqref{freniamolentropia} and \eqref{freniamolenormali} it is immediate to show that  $\mathcal{U}=T^{00}$ can be written as the Legendre transform of $\Lambda$ with respect to spatial part of the superfluid currents,
\begin{equation}\label{dubbidubadubbidubbiduba}
\mathcal{U} = -\Lambda + \sum_{i=1}^k \mu^i_j n^j_i \, .
\end{equation}
This implies that the natural set variables upon which $\mathcal{U}$ depends is 
\begin{equation}\label{chartofU}
(n_A,n_i^0,\mu_j^i)^{A=0,...,l}_{i=1,...,k}
\end{equation}
and its differential, employing the definitions \eqref{temperiamolatemperatura2} and \eqref{chemichiamolachimica}, reads
\begin{equation}\label{ilcentroditutto}
d\mathcal{U} =\Theta ds + \sum_{A=1}^l \mu_A^{(T)} dn_A + \sum_{i=1}^k \mu_i^{(T)} dn_i^0 + \sum_{i=1}^k  n_i^j d\mu^i_j.
\end{equation}
This formula alone contains the essence of all the results we collected up to now: since we can associate to any superfluid current a conjugate varibale that is conserved during the equilibration process (the momentum), at a thermodynamic level we need to include a dependence on it in the equation of state. 
Moreover, the conservation of the superfluid momentum is the reason why, to get the temperature $\Theta$ or the thermodynamic chemical potentials $\mu_x^{(T)}$, the superfluid momenta must be fixed when performing the partial derivatives of $\mathcal{U}$.

Now that we identified the internal energy and its natural set of variables, we can make further Legendre transformations to define other thermodynamic potentials. If we transform with respect to the entropy density we get the Helmholtz free energy density
\begin{equation}
\mathcal{F} = \mathcal{U} - \Theta s ,
\end{equation}
whose differential (on the manifold $\mathcal{Q}$) is
\begin{equation}
d\mathcal{F} = -s d\Theta + \sum_{A=1}^l \mu_A^{(T)} dn_A + \sum_{i=1}^k \mu_i^{(T)} dn_i^0 + \sum_{i=1}^k  n_i^j d\mu^i_j.
\end{equation}
Notice that the function $\mathcal{F}$ defined in \eqref{energialiberallinizio} reduces to the Helmholtz free energy when the equilibrium is reached, which justifies the use of the same name.

We can also introduce the grand potential density
\begin{equation}
\mathcal{K} = \mathcal{U}-\sum_x n_x^0 \mu^{x0} \, , 
\label{grandpotential}
\end{equation}
which, considering equations \eqref{dubbidubadubbidubbiduba} and \eqref{pressionciona}, is found to be equal to $-\Psi$. Its differential is
\begin{equation}
d \mathcal{K}  = -s d \Theta - \sum_{A=1}^l n_{A} d\mu_{A}^{(T)}  - \sum_{i=1}^k n_{i}^0 d\mu_{i}^{(T)}  + \sum_{i=1}^k n_{i}^j d\mu_{j}^i \, .
\end{equation}
If in the above equation we substitute $\mathcal{K}$ with $-\Psi$ we get the generalization of the Gibbs-Duhem  relation for multifluids. 

The presence of superfluid components extends the set of possible thermodynamic potentials. For example, the Legendre transform of the internal energy with respect to all the superfluid momenta reads
\begin{equation}\label{EEEEEEEEEEEEEEEEEEEEEEEEEEE}
\mathcal{E} = \mathcal{U} - \sum_{i=1}^k n_i^j \mu^i_j ,
\end{equation}  
whose differential is
\begin{equation}
d\mathcal{E} =\Theta ds + \sum_{A=1}^l \mu_A^{(T)} dn_A + \sum_{i=1}^k \mu_i^{(T)} dn_i^0 - \sum_{i=1}^k   \mu^i_j dn_i^j.
\end{equation}
Comparing equations \eqref{dubbidubadubbidubbiduba} and \eqref{EEEEEEEEEEEEEEEEEEEEEEEEEEE} we obtain
\begin{equation}\label{redunctionism}
\mathcal{E} = -\Lambda \, .
\end{equation}
Despite its simplicity, this result plays a central role in our thermodynamic analysis, as it unveils the thermodynamic meaning of the master function: it is obtainable by means of a Legendre transformation of the internal energy density $\mathcal{U}$ with respect to the superfluid momenta. From the practical point of view, this procedure is important because it can be used to reconstruct the master function directly from an equation of state for the internal energy (or from any other thermodynamic potential supplied by a microphysical calculation).

For the sake of completeness we present a further thermodynamic potential, 
\begin{equation}
\mathcal{J}= \mathcal{U}-\sum_{i=1}^k n_i^0 \mu^{i0},
\end{equation}
whose thermodynamic differential is
\begin{equation}
d\mathcal{J} =\Theta ds + \sum_{A=1}^l \mu_A^{(T)} dn_A - \sum_{i=1}^k n_i^0 d \mu_i^{(T)} + \sum_{i=1}^k  n_i^j d\mu^i_j.
\end{equation}
The potential $\mathcal{J}$ depends on an interesting set of state variables. In microphysical calculations, to define the phases $\phi^i$ of the order parameters, one is usually forced to break the conservation of the number of particles of the superfluid species \citep{anderson_considerations}. 
This suggests  to work with an ensemble in which the chemical potential $\mu_i^{(T)}$ (which by Josephson relation \eqref{topologiallariscossa} is proportional to $\partial_t \phi^i$) is assigned and $n_i^0$ is computed as a quantum-statistical average. For this reason $\mathcal{J}$ is a natural alternative to $\mathcal{U}$ in the construction of an equation of state.

\subsection{The Euler relation}

Like in the thermodynamic study of a simple fluid, it is possible to derive the Euler (or fundamental) relation also in the multifluid case \citep[see e.g. the discussion in ][]{andersson2007review}. Since the Euler relation stems from the additivity property of the system, its derivation is typically performed by employing the thermodynamic potential which is a natural function of all the extensive quantities. 
In our case this thermodynamic potential is $\mathcal{E}$ and equation \eqref{pressionciona} implies that the fundamental relation is
\begin{equation}
\label{zerbino}
\mathcal{E} = -\Psi + \sum_x n_x^0 \mu^{x0} - \sum_{i=1}^k n_{i}^j\mu_{ j}^i \, .
\end{equation}
It is interesting to recast the fundamental relation in terms of the internal energy:
\begin{equation}
\label{euler2}
\mathcal{U} = -\Psi + \Theta s + \sum_{A=1}^l n_{A} \mu_{A}^{(T)} + \sum_{i=1}^k n_{i}^0 \mu_{i}^{(T)}.  
\end{equation}
Since it does not explicitly contain any term  associated with the presence of superfluids in the right-hand side, this form of the Euler relation is formally identical to the one for a mixture of normal fluids. This is due to the fact that the set of natural variables of $\mathcal{U}$, presented in \eqref{chartofU}, contains the spatial components of the momenta per particle, which are intrinsically intensive quantities and, therefore, cannot produce the associated terms in the Euler relation.

It is possible to write the fundamental relation \eqref{zerbino} also in a manifestly covariant way that is, however, less physically transparent. For example, in \cite{prix2005}, the name  ``chemical potentials'' is assigned to the quantities
\begin{equation}\label{comcomcomcomcocm}
\mu_x^{(C)} = -\mu^x_\nu u_x^\nu \, ,
\end{equation}
where $u_x^\nu$ is the four-velocity associated to the species $x$. We use the label $C$, here, which stands for \textit{comoving}, to distinguish them from those defined in \eqref{chemichiamolachimica}, in which we use a $T$, for \textit{thermodynamic}. According to this definition, the Euler relation becomes \citep[see also ][]{rezzolla_book}
\begin{equation}\label{euler1}
\mathcal{E} = -\Psi + \Theta s + \sum_{x \neq s} n_x \mu_x^{(C)} \, . 
\end{equation} 
Notice that \eqref{euler1} and \eqref{euler2} are both very similar to the usual Euler relation for a simple fluid. However, one must not be tempted to identify $\mathcal{E}$ with the thermodynamic energy density as this would lead to the confusion between the thermodynamic and the comoving chemical potentials, which can assume very different values. An interpretation problem related to the different chemical potentials used in the two-fluid models of \cite{pca2002} and \cite{KP2017} is discussed and solved in appendix \ref{kasumotomorizuma}.

\subsection{Entrainment and thermodynamics}\label{entrainment and thermodyanmics}

One of the properties of a multifluid is the possibility to have non-collinearity between the currents and their conjugate momenta, which gives rise to the so-called entrainment effect. In the formulation presented in section \ref{sec:Gftva} the momenta naturally arise as linear combinations of the currents, see \eqref{noivinciamosempre}:
\begin{equation}\label{entroinentrainmentconte}
\mu_\nu^x = \sum_y \mathcal{B}^{xy} n_{y\nu},
\end{equation}
where we have introduced the symmetric matrix $\mathcal{B}^{xy}$ such that
\begin{equation}
\mathcal{B}^{xx} := \mathcal{B}^x  \spc  \mathcal{B}^{xy} := \mathcal{A}^{xy}
\end{equation}
for $x\neq y$. On the other hand, if we define the symmetric matrix $\mathcal{B}_{xy}$ to be the inverse of $\mathcal{B}^{xy}$,
\begin{equation}
\sum_y \mathcal{B}^{xy}\mathcal{B}_{yz} = \delta\indices{^x _z},
\end{equation}
we can write the currents as linear combinations of the momenta:
\begin{equation}
\label{invertito!!!}
n_x^\nu = \sum_y \mathcal{B}_{xy}\mu^{y\nu}.
\end{equation}
This relationship between momenta and currents is the most natural when we start from the momentum-based ``potential variational principle''  instead of the ``current-based'' one of section \ref{sec:Gftva}. 
As pointed out by \cite{CarterKhalat92}, see also  \cite{Prix_single_vortex} and \cite{Schmitt_Instabilities}, in the potential variational principle the master function of the theory is $\Psi = -\mathcal{K}$ written as a function of the scalar products $ \mu_{xy}^2 = -\mu^x_\nu \mu^{y\nu} $, in place of $\Lambda=-\mathcal{E}$ written as a function of $n_{xy}^2$.

In a thermodynamic perspective, we are interested in writing the normal momenta and the superfluid currents as linear combinations of the normal currents and of the superfluid momenta. The reason for this is that the normal currents identify the reference frame in which the thermodynamic quantities must be computed, while the superfluid momenta enter directly into the equation of state. 
This representation of the entrainment arises naturally from a hybrid variational principle which uses $\mathcal{X}=-\mathcal{J}$ as the master function, see e.g. \cite{CarterKhalat92}, \cite{cool1995} and \cite{Dubovsky2012}. 
The entrainment matrix for this hybrid description can be written with a little algebra in terms of $\mathcal{B}_{xy}$, as pointed out in \cite{Prix_single_vortex}. It is now important to briefly review and extend this set of ideas. 

To introduce the hybrid representation, let us separate  the normal components from the superfluid ones in equation \eqref{invertito!!!}, namely
\begin{equation}
\begin{split}
& n_A^\nu = \sum_{B=0}^l \mathcal{V}_{AB} \mu^{B\nu} + \sum_{h=1}^k \mathcal{M}_{Ah} \mu^{h\nu} \\
& n_i^\nu = \sum_{B=0}^l \mathcal{M}_{Bi} \mu^{B\nu} + \sum_{h=1}^k \mathcal{S}_{ih} \mu^{h\nu},\\
\end{split}
\end{equation}
where $A=0$ stands for $s$. The three matrices appearing in the right hand side are respectively the $(1+l)\times (1+l)$ symmetric matrix
\begin{equation}
\mathcal{V}_{AB} = \mathcal{B}_{AB},
\end{equation}
the $k \times k$ symmetric matrix
\begin{equation}
\mathcal{S}_{ih} = \mathcal{B}_{ih}
\end{equation}
and the $(1+l)\times k$ matrix
\begin{equation}
\mathcal{M}_{Ai} = \mathcal{B}_{Ai}.
\end{equation}
With a little algebra it is possible to show that
\begin{equation}
\label{ancoranacoravorrei}
\begin{split}
& \mu^A_\nu = \sum_{B=0}^l (\mathcal{V}^{-1})^{AB} n_{B\nu} - \sum_{h=1}^k \mathcal{D}\indices{^A _h} \mu^{h}_\nu \\
& n_i^\nu = \sum_{B=0}^l \mathcal{D}\indices{^B _i} n_B^\nu + \sum_{h=1}^k Y_{ih} \mu^{h\nu},\\
\end{split}
\end{equation}
where we have introduced the $(1+l)\times k$ matrix
\begin{equation}
 \mathcal{D}\indices{^A _i} = \sum_{B=0}^l (\mathcal{V}^{-1})^{AB} \mathcal{M}_{Bi} 
\end{equation}
and the $k \times k$ symmetric matrix
\begin{equation}
Y_{ih} = \mathcal{S}_{ih} - \sum_{B,C =0}^l (\mathcal{V}^{-1})^{BC} \mathcal{M}_{Bi} \mathcal{M}_{Ch}.
\end{equation}
With calculations analogous to those made to obtain the system \eqref{ancoranacoravorrei}, the superfluid block of the entrainment matrix given in \eqref{entroinentrainmentconte} is 
\begin{equation}\label{uscendoentrainmentconte}
\mathcal{B}^{ih} = (Y^{-1})^{ih}.
\end{equation} 
The matrix $Y_{ih}$ naturally encodes the notion of entrainment in a thermodynamic context. To give a precise meaning to this sentence, let us consider again equation \eqref{ilcentroditutto}. Combining \eqref{ancoranacoravorrei}, \eqref{freniamolentropia} and \eqref{freniamolenormali} we get
\begin{equation}
n_i^j = \sum_{h=1}^k Y_{ih} \mu^{hj},
\end{equation}
which, using the symmetry of $Y_{ih}$, implies
\begin{equation}\label{ilcentroditutto2}
d\mathcal{U} =\Theta ds + \sum_{A=1}^l \mu_A^{(T)} dn_A + \sum_{i=1}^k \mu_i^{(T)} dn_i^0 + \sum_{i,h=1}^k \dfrac{Y_{ih}}{2}  d(\mu^{hj}\mu^i_j).
\end{equation}
This tells us that, in a thermodynamic perspective, the interesting part of the entrainment matrix is the one that defines the relationship between the superfluid momenta and the superfluid currents in the reference frame of the heat bath. In fact, in the following section we  show that $Y_{ih}$ is the only physically important part of the entrainment matrix in a non-dissipative theory, while $\mathcal{V}_{AB}$ and $\mathcal{D}\indices{^A _i}$ have a microscopic interpretation only if one takes into account dissipative effects, so their knowledge is not needed and cannot be uniquely extracted within the framework of equilibrium thermodynamics. 

\section{Local thermodynamics of perfect multifluids}
\label{Local thermodynamics of perfect multifluids}

In this section we seek for a formulation of multifluid hydrodynamics in which the  thermodynamic interpretation of the variables is manifest. This is certainly possible for a perfect multifluid since, in the absence of dissipation, the second principle ensures that the system is in local thermodynamic equilibrium. Hence, the  formalism developed in section \ref{Equilibrium properties of homogeneous systems} should be locally valid in a frame comoving with the entropy.

Before getting into the details, let us comment on how the equations of motion of the phenomenological model of section \eqref{variationalprinciplesubsection} are modified in view of the results of the previous section.  Clearly the  normal currents must be implemented as s-locked currents, while the superfluid ones  as free. Moreover, superfluidity imposes the irrotationality of the momenta. Hence, the full system of hydrodynamic equations for a perfect multifluid is
\begin{equation}\label{dynamicsss}
\begin{split}
& \nabla_\nu n_x^\nu=0 \\
& \partial_{[\rho} \mu^i_{\nu]}=0 \\
& \nabla_\nu T\indices{^\nu _\rho}=0. \\
\end{split}
\end{equation} 
This system is compatible with  equations \eqref{equationsofmotioningeneral}, the only difference is that the irrotationality requirement adds a further constraint to the initial condition.

\subsection{The local frame of thermodynamics}
\label{The local frame of thermodynamics}

It is instructive to rework the Lagrangian approach of section \ref{sec:Gftva}  with the aid of a tetrad comoving with $s^\mu$, i.e. a collection of four vector fields $e_a=e^\nu _a \partial_\nu$, $a=0,1,2,3$ that are orthonormal\footnote{
The fields $e_a$ are assumed to be linearly independent everywhere, so that we can define the matrix $e^a_\nu$ to be the inverse of $e_a^\nu$, which means $e^a_\nu e^\nu_b = \delta\indices{^a _b}$ and $e^\nu_a e^a_\rho = \delta\indices{^\nu _\rho}$. Moreover, they satisfy the orthonormality condition 
$g(e_a,e_b) = \eta_{ab}$. In the following, for the tetrad indices we will use the convention that the generic latin indices $a$ and $b$ run over all the possible indices (from $0$ to $3$), while $j=1,2,3$ are restricted to the spatial ones, in accordance with the convention introduced in the previous section. Einstein summation convention will be applied to both $a$ and $j$ indices.
}
and such that $e_0 = u$. This tetrad represents the local Minkowskian frame of an observer $\mathcal{O}$ moving with four-velocity $u$. The components of a generic tensor on this basis can be interpreted as the physical quantities seen by $\mathcal{O}$ \citep[see e.g. ][]{special_in_gen}. 

The physical motivation for considering this particular tetrad is that the normal currents observed by $\mathcal{O}$ are $n_A^j=0$, for $A=0,...,l$. Therefore, a perfect multifluid is \emph{locally} in a state belonging to $\mathcal{Q}$, the manifold of equilibrium macrostates defined in section \ref{thermopetentissimi}. This implies that the energy density measured by $\mathcal{O}$ coincides with the thermodynamic variable $\mathcal{U}$, namely
\begin{equation}\label{interiormente}
u_\nu u_\rho T^{\nu \rho} = \mathcal{U} .
\end{equation}
Now, equations \eqref{temperiamolatemperatura2} and \eqref{chemichiamolachimica} define the local temperature and chemical potentials of the multifluid; in the tetrad formalism they simply become
\begin{equation}
\Theta = \Theta^0 \spc \spc \mu_x^{(T)} = \mu^{x0}.
\end{equation}
For later convenience we also introduce the two orthogonal projectors
\begin{equation}
\label{proettoriferoci1}
\paral\indices{^\nu _\rho}:= - u^\nu u_\rho = e^\nu_0 e^0_\rho
\qquad
\perp\indices{^\nu _\rho}:= \delta\indices{^\nu_\rho} + u^\nu u_\rho =  e^\nu_j e^j_\rho\, . 
\end{equation} 
Since these two projectors are complementary, i.e.
\begin{equation}\label{resolutionofidentity}
\delta\indices{^\nu _\rho} = \paral\indices{^\nu _\rho} + \perp\indices{^\nu _\rho}  \, ,
\end{equation}
they can be used to separate the time and the space parts of a vector in the tetrad frame: the tensor $\paral\indices{^\nu _\rho}$ projects any vector along $u$, while $\perp\indices{^\nu _\rho}$ projects it 
tangentially to the local present of the observer $\mathcal{O}$.

\subsection{The emergence of the Landau-Andreev-Bashkin formalism}
\label{bsgformalism}

We now use the tetrad to endow the non-homogeneous phenomenological model with a  language that is more directly in contact with the microphysics of the system. In pursuing this aim, we find that such a language is the one proposed by \cite{GusakovAndersson2006}, which, in turn, is the multi-component generalization of the single-component  superfluid model of \cite{Son2001}.
While the work of Son provides a natural relativistic extension of the non-relativistic two-fluid model pioneered by Tisza and Landau \citep[see e.g. ][]{landau6}, the formalism developed by Gusakov can be considered as the natural formulation in General Relativity of the multifluid hydrodynamics pionereed by \cite{andreevbashkin1976}.
For this reason, in the following, we will speak of relativistic Landau-Andreev-Bashkin (LAB) formalism, as opposed to the phenomenological Carter formalism of section  \ref{sec:Gftva}.
%
%
Here we show how the LAB formalism naturally emerges from the Carter's approach in view of the thermodynamic considerations of section \ref{Equilibrium properties of homogeneous systems}.

Let us define the \textit{winding vector}
\footnote{
Following Landau's original terminology, in \cite{GusakovAndersson2006} and \cite{Gusakov2007} the quantity $w_i^\nu$ is referred to as a ``four-velocity'' for the motion of a superfluid component. Since $w_i^\nu$ is not normalized to $-1$, we preferer to call it ``winding vector'' in view of equation \eqref{quantizzo per sempre}. } 
to be the spatial part of the superfluid momentum relative to a superfluid species $i$, namely
\begin{equation}\label{defhydro1}
w_{i\nu} := \perp\indices{^\rho _\nu} \mu^i_\rho = \mu^i_j e^j_\nu  \, .
\end{equation}
With the aid of \eqref{resolutionofidentity}, the superfluid momenta $\mu^i_\nu$  are split as
\begin{equation}\label{ilcatalicammello3}
\mu^i_\nu = \mu_i^{(T)} u_\nu + w_{i\nu} .
\end{equation}
On the other hand,  the same operation performed on the superfluid currents $n_i^\nu$ gives
\begin{equation}\label{ilcatalicammello5}
n_i^\nu = n_i^{(T)} u^\nu + \sum_h Y_{ih} w^{\nu}_h,
\end{equation} 
where we have employed equation \eqref{ancoranacoravorrei} and the definition
\begin{equation}\label{defhydro2}
n_i^{(T)} := -u_\nu n_i^\nu,
\end{equation}
which is the the density of the species $i$  measured by $\mathcal{O}$.

Finally, we split the energy-momentum tensor as
\begin{equation}
\begin{split}
T^{\nu \rho}& =  T^{ab}e_a^\nu e_b^\rho  \\
& = T^{00}e_0^\nu e_0^\rho + T^{j0}(e_0^\nu e_j^\rho + e_j^\nu e_0^\rho) + T^{jj'}e_j^\nu e_{j'}^\rho.
\end{split}
\end{equation}
Thanks to equation \eqref{interiormente} and the fact that  $n_A^j =0$ in the tetrad frame, we can identify the various components of $T^{ab}$ as
\begin{equation}
\begin{split}
& T^{00} = \mathcal{U} \\
& T^{j0} =\sum_{i=1}^k n_{i}^j \mu^{i0} \\
& T^{jj'} = \Psi \eta^{jj'} + \sum_{i=1}^k n_{i}^j \mu^{ij'} ,
\end{split}
\end{equation}
which leads to
\begin{equation}
\label{dameAAA}
\begin{split}
 T^{\nu \rho} &= (\mathcal{U} + \Psi) u^\nu u^\rho + \Psi g^{\nu \rho}   + \\
& + \sum_{i,h=1}^k Y_{ih} (\mu_{i}^{(T)} u^\nu w^{\rho}_h + \mu_{h}^{(T)} u^\rho w^{\nu}_i + w^{\nu}_i w^{\rho}_h).
\end{split}
\end{equation}
Interestingly, this reworking of the energy momentum tensor makes evident that a mixture of normal components (i.e. when $k=0$) is equivalent to a perfect fluid.
We now can combine \eqref{ilcentroditutto2} with \eqref{defhydro1} and \eqref{defhydro2} to get an expression for the differential of $\mathcal{U}$ that involves only variables that are manifestly covariant, 
\begin{equation}\label{ilcentroditutto3}
d\mathcal{U} =\Theta ds + \sum_{A=1}^l \mu_A^{(T)} dn_A + \sum_{i=1}^k \mu_i^{(T)} dn_i^{(T)} + \sum_{i,h=1}^k \dfrac{Y_{ih}}{2}  d(w_{i}^\nu w_{h\nu}),
\end{equation}
One can verify that  \eqref{u_T}, \eqref{s-locking}, \eqref{ilcatalicammello5}, \eqref{ilcatalicammello3},  \eqref{dameAAA}, \eqref{ilcentroditutto3} and \eqref{dynamicsss} define a complete (and also redundant) self-consistent system of hydrodynamic equations. 
Hence, for a generic perfect multifluid the LAB formalism is completely equivalent to the Lagrangian approach of Carter. 
In particular, the equivalence between the Son's relativistic version of the Landau two-fluid model and the Carter's formalism has already been proven in \cite{andersson2007review}; here we generalised the discussion by considering the most general theory for a perfect multifluid and we provided a dictionary to easily translate the two descriptions one into the other. 

%

\subsection{The problem of the redundancy of the master function}

We casted the perfect multifluid hydrodynamics in a way that only the knowledge of  $\mathcal{U}$ (or any other thermodynamic potential) is required. However, given the expression for $\mathcal{U}$, one would like to also have a technique to write the corresponding master function. 

Since $\Lambda=-\mathcal{E}$, the first obvious step is to obtain $\mathcal{E}$ from $\mathcal{U}$ via the Legendre transform \eqref{EEEEEEEEEEEEEEEEEEEEEEEEEEE}. However, this is not sufficient to complete the task:  $\Lambda$ is defined as a function of the scalars $n^2_{xy}$, while by making the Legendre transform of $\mathcal{U}$ we arrive at $\mathcal{E}=\mathcal{E}(n_x^0,n_i^j)$. 
Rewriting the dependences correctly may seem straightforward but there is a subtlety which arises from the fact that some of the components are s-locked. Before going into the details it is convenient to give an idea of where the complication comes from. 

Consider a simple perfect fluid at finite temperature, i.e. $l=1$ and $k=0$. 
The two currents of the theory are
\begin{equation}\label{1234}
s^\nu = s u^\nu  \spc  n^\nu = n u^\nu,
\end{equation}  
where $n^\nu$ is the (s-locked) current of particles. The energy-momentum tensor can be written accoprding to the  the LAB formalism by setting $k=0$ into \eqref{dameAAA},
\begin{equation}\label{5678}
T^{\nu \rho} = (\mathcal{U} + \Psi) u^\nu u^\rho + \Psi g^{\nu \rho}.
\end{equation}
According to the LAB formulation, the only microscopic input needed to describe a perfect fluid is an equation of state of the form
\begin{equation}
\mathcal{U} = \mathcal{U}(s,n),
\end{equation}
whose differential is
\begin{equation}
d\mathcal{U} = \Theta ds + \mu^{(T)}dn.
\end{equation}
Now, let us consider the master function 
\begin{equation}
\Lambda = \Lambda (n_{ss}^2,n_{sn}^2,n_{nn}^2).
\end{equation}
Combining \eqref{dubbidubadubbidubbiduba}, \eqref{freniamolentropia} and \eqref{freniamolenormali}, it is immediate to find that
\begin{equation}
\mathcal{U}(s,n) = -\Lambda (s^2,sn,n^2).
\end{equation}
Therefore, to construct a function $\mathcal{U}$ that fully describes the fluid, only the behaviour of $\Lambda$ on the surface
\begin{equation}\label{9012}
n_{sn}^2 = \sqrt{n_{ss}^2 n_{nn}^2},
\end{equation} 
is needed. On the other hand, the values of $\Lambda$ elsewhere constitute a part of information which is completely lost when the Son and Gusakov formulation of hydrodynamics is used. In other words, two different master functions $\Lambda$ and $\Pi$ which are equal on this surface would produce the same LAB description and, therefore, the same physical theory. 
This redundancy in the arguments of the master function arises because of the collinearity requirement for $s^\nu$ and $n^\nu$: the two functions $\Lambda$ and $\Pi$ are equal on all the possible equilibrium macrostates.

The fact that $\Lambda$ and $\Pi$ describe the same physical system implies that there is not a unique prescription to define $\mathcal{V}_{AB}$: there are many equivalent ways to construct the entrainment matrix.
This arbitrariness  is a peculiarity of the non-dissipative theory. In fact, the behaviour of $\Lambda$ on the macrostates in which $s^\nu$ and $n^\nu$ are not locked becomes relevant for dissipative models with heat transport. In fact, for dissipative systems the entrainment with the entropy component becomes a necessary and physically meaningful element of the description \citep[see e.g. ][]{lopez2011}. On the contrary, in the non-dissipative limit this arbitrariness may be advantageous as it allows to set the entrainment anomaly $\mathcal{A}^{sn}$ to zero imposing
\begin{equation}
\Lambda=-\mathcal{U}(\sqrt{n_{ss}^2}, \sqrt{n_{nn}^2}).
\end{equation}
In this simple case with $k=0$ it does not seem to make any difference the fact that the entrainment matrix is not uniquely defined: since there are no superfluid components, the currents are all parallel with respect to each other and any master function we choose will produce in the end the same momenta
\begin{equation}
\Theta_\nu = \Theta u_\nu \spc \mu_\nu = \mu^{(T)}u_\nu .
\end{equation}
%
However, in the case of a generic perfect multifluid, not only $\mathcal{V}_{AB}$ but also $\mathcal{D}\indices{^A _i}$ is not uniquely defined; a change of the master function can modify the definition of the normal momenta. 
The following three subsections are be devoted to the analysis of this non-uniqueness of the normal momenta. In section \ref{Mathematical explanation of the redundancy} we study the redundancy problem of the master function starting from the action principle, in \ref{Thermodynamic origin of the redundancy} we  analyse its physical interpretation in a thermodynamic perspective. Finally, in \ref{A concrete example} we  provide a practical example in which we show how to fix the master function to remove the entrainment associated to the entropy in neutron star matter.

\subsection{Mathematical explanation of the redundancy}
\label{Mathematical explanation of the redundancy}

Let us consider, for a generic multifluid, two different functions $\Lambda$ and $\Pi$ from $\mathbb{R}^z$ to $\mathbb{R}$. The $z$ real arguments of these functions are the numbers $n_{xy}^2$, considered as independent variables. If the normal species are forced to comove with the entropy, i.e. when
\begin{equation}
n_A^\nu = n_A \dfrac{s^\nu}{s},
\end{equation}
the arguments of $\Lambda$ and $\Pi$ are no more free, but satisfy the conditions
\begin{equation}
n_{Ax}^2 = \dfrac{n_A}{s} n_{sx}^2 .
\end{equation}
Assume that $\Lambda$ and $\Pi$ are equal when the above constraints are satisfied but that are different on an out of local equilibrium macrostate of the multifluid. Remembering that the normal currents comove with the entropy both on and off shell, according to equation \eqref{ilbuglagrangiano} we have
\begin{equation}
\begin{split}
\mathcal{L}[ g  , X_s^\alpha  ,X^\alpha_{i} ]& =  \Lambda (-g_{\rho \nu} \; n_x^\rho[ g  , X_s^\alpha  ,X^\alpha_{i} ] \; n_y^\nu[ g  , X_s^\alpha  ,X^\alpha_{i} ] )  \\
& \equiv \Pi (-g_{\rho \nu} \; n_x^\rho[ g  , X_s^\alpha  ,X^\alpha_{i} ] \; n_y^\nu[ g  , X_s^\alpha  ,X^\alpha_{i} ] ). \\
\end{split}
\end{equation}
This means that, since the energy-momentum tensor and all the hydrodynamic equations are calculated only imposing the stationarity of the action, the choice between $\Lambda$ and $\Pi$ does not affect the physics of the perfect multifluid, in the sense that they give rise to the same physical theory: both can be used to construct the same equations of motion and energy-momentum tensor, which can then be used to uniquely derive the LAB description of the system by using the method of section \eqref{bsgformalism}.

However, the choice between $\Lambda$ and $\Pi$ can modify concretely the normal momenta. To see this, consider a perfect multifluid on shell, choose a point of the spacetime and compute the component $\mu_{ j}^A$ in the basis inrtoduced in section \ref{The local frame of thermodynamics}, for  $A \neq s$, directly from the master function. By definition $n_A^j=0$, so that
\begin{equation}
\mu_{ j}^A = \dfrac{\partial \Lambda}{\partial n_{A}^j} \bigg|_{n_A^j=0} 
= 
\lim_{h \rightarrow 0} \dfrac{\Lambda(n_{A}^j =h) -\Lambda(n_{A}^j=0 )}{h} \, ,
\end{equation}
where all the remaining $n_x^a$ are kept fixed. Since the derivative is computed for $s^j=0$, the quantity $\Lambda(n_{A}^j =h)$ is the master function evaluated on a macrostate in which the currents $n_{A}^\nu$ and $s^\nu$ are not parallel with respect to each other: performing the above derivative implies that we are exploring a domain in which $\Lambda$ may differ from $\Pi$. 

The freedom to choose the master function on the mascrostates in which one (or more) s-locked current does not comove with the entropy can be seen as a gauge freedom of the non-dissipative theory. Only performing a gauge fixing (i.e. assigning the master function in a neighbourhood of the equilibrium) it is possible to define uniquely the normal momenta. Notice that only the spatial part $ \perp\indices{^\rho_\nu} \mu_{\rho}^A$ of the normal momenta is affected by a gauge-fixing of the master function. In fact, the previous argument is no more valid for $\mu^A_0$, because the variation of the argument is performed in the direction of the entropy:
\begin{equation}
\mu_{ 0}^A = \dfrac{\partial \Lambda}{\partial n_{A}^0}  
= 
\lim_{h \rightarrow 0} \dfrac{\Lambda(n_A^j=0, n_{A}^0 + h) -\Lambda(n_A^j=0, n_{A}^0)}{h} \, .
\end{equation}
Also the superfluid momenta are uniquely determined, because in the corresponding derivation we have to move the superfluid current keeping the normal ones collinear. This justifies the statement, made in section \ref{entrainment and thermodyanmics}, that in a non-dissipative theory $Y_{ih}$ is the only physically meaningful part of the entrainment matrix. Stated in other words, according to equation \eqref{uscendoentrainmentconte} only the block $\mathcal{B}^{ih}$ of the full entrainment matrix $\mathcal{B}^{xy}$ is gauge-independent.

Since the choice between $\Lambda$ and $\Pi$ does not affect the energy-momentum tensor, the transformations $\mu^A_j \longrightarrow \tilde{\mu}^A_j$ produced in the passage from $\Lambda$ to $\Pi$, are not completely free, but must satisfy some constrains. In fact, the currents produced by considering the two different gauge-fixing must satisfy
\begin{equation}
s^0 \Theta_j + \sum_{A=1}^l n_{A}^0 \mu_{ j}^A + = s^0 \tilde{\Theta}_j + \sum_{A=1}^l n_{A}^0 \tilde{\mu}_{ j}^A \, .  
\end{equation}
This can be proven by considering the general expression of $T^{\nu \rho}$ in  \eqref{energiamomentotempfinita}, together with the fact that the energy-momentum tensor (in particular $T\indices{^0 _j}$) and the superfluid momenta are gauge-invariant quantities.
We can use the above constraint to make explicit what happens when a gauge transformation is performed:
\begin{equation} 
\label{gaugeingenerale}
\begin{split}
&\tilde{\mu}_{ \nu}^i = \mu_{ \nu}^i
\\
&\tilde{\mu}_{ \nu}^A = \mu_{\nu}^A +  \pi_{ \nu}^A
\\
&\tilde{\Theta}_{\nu} = \Theta_\nu + \zeta_\nu \, ,  
\end{split}
\end{equation}
where
\begin{equation}
\label{bazinga}
s^\nu \pi_{\nu}^A =0 
\qquad \qquad
\zeta_\nu = - \dfrac{1}{s} \sum_{A=1}^l n_{A} \pi_{\nu}^A.
\end{equation}
Therefore, the  gauge transformation arising from a different choice of the master function amounts to  a redistribution of the spatial part of the momentum between the normal currents.

Finally, let us remark that the gauge-freedom discussed here should not be confused with the notion of \emph{chemical gauge} introduced in \cite{carter_macro_2006}, as discussed in appendix \ref{Chemical gauge in an arbitrary perfect multifluid}.

\subsection{Gauge-invariant quantities}
\label{Thermodynamic origin of the redundancy}

The fact that two different master functions (say the $\Lambda$ and $\Pi$ of the previous subsection) may produce the same LAB description of a given multifluid system can be used to understand in a physical perspective why some quantities are invariant under the gauge transformation \eqref{gaugeingenerale}. Let us examine them one by one.

According to \eqref{gaugeingenerale}, the superfluid momenta are not affected by a gauge fixing of the master function. It is therefore possible to interpret them as gradients of the  phases $\phi^i$ 
(i.e. to assign them a physical interpretation that is beyond the Carter's phenomenological approach).


Now, consider equations  \eqref{dameAAA} and \eqref{ilcentroditutto3}, which define the natural variables of the LAB description. Equation  \eqref{gaugeingenerale} tells us that the differentiated variables (i.e. $\mu_A^{(T)}$, $\mu_i^{(T)}$ and $w_i^{\nu}$) in the right-hand side of \eqref{ilcentroditutto3} are gauge-invariant: 
since $\mathcal{U}$ is gauge invariant, so are also all its derivatives, namely $\Theta=-\Theta_\nu u^\nu$, $\mu_x^{(T)}=-\mu^x_\nu u^\nu$ and $Y_{ih}$. In particular, the fact that $\Theta$ and $\mu_x^{(T)}$ are gauge-invariant further justifies their interpretation as the absolute temperature and the chemical potentials of the system.

Also the generalized pressure $\Psi = -\mathcal{K}$ is a gauge invariant, as it can be obtained from $\mathcal{U}$ via Legendre transform.
In brief, all the quantities that naturally appear in the LAB description must be gauge-invariant (i.e. are unaffected by a gauge fixing of the master function). 

Finally, the normal currents do not explicitly appear in \eqref{ilcentroditutto3}. Even if equation \eqref{energiamomentotempfinita} seems to suggest that the knowledge of $\mu^A_j$ is required to construct the energy-momentum tensor, there is no need to specify them, due to the fact that the symmerty of the energy-momentum tensor $T^{0j}=T^{j0}$ yields
\begin{equation}
\sum_{A=0}^l n_A^0 \mu^A_j + \sum_{i=1}^k n_i^0 \mu^i_j = \sum_{i=1}^k \mu^{i0} n_{ij}.
\end{equation}
This proves that the only part of the normal momenta whose knowledge is explicitly required is the thermodynamic chemical potential. 

To give a physical interpretation of this kind of gauge freedom we may start interpreting $\mu_j^A$ as the average momentum per particle of the species $A$. Now, without a microphysical insight it is difficult, in equilibrium thermodynamics, to separate the contribution which comes from the elementary excitations and should strictly contribute to $\Theta_j$ from the genuine momentum of the species $A$. 
%
%
This ambiguity describes the absence of a entirely \emph{macroscopic} criterion to distribute the momentum between the normal components in the non-dissipative limit of the theory: the only prescription to uniquely assign the entrainment matrix is to consider a dissipative theory in which the normal currents are not forced to comove with the entropy. In this case, the heat contribution to the energy-momentum tensor contributes to fix the entrainment matrix \citep[see e.g.][]{lopez2011}. 

On the other hand, if the purpose is to construct a theory for a given perfect multifluid, it is possible to take advantage of this fundamental gauge freedom to choose a master function that is particularly convenient for the specific system under study.

\subsection{A practical example of gauge fixing}
\label{A concrete example}

The minimal model of superfluid neutron stars consists of two distinct interpenetrating dynamical components, one normal $(l=1)$ and one superfluid $(k=1)$, see e.g. \cite{chamel_review_crust}. Following \cite{langlois98}, the superfluid current $n_n^\nu$ is comprised of free neutrons, while the normal current  $n_p^\nu$  can be defined as the remaining part of the conserved total baryon current. In this subsection we exploit the gauge freedom to show that there is no loss of generality in tuning such a minimal model in a way that the entropy does not carry any entrainment, namely
\begin{equation}\label{carterizzoilthermal}
{\perp}\indices{^\rho _\nu} \Theta_\rho =0.
\end{equation}
Consider the generic master function 
\begin{equation}
\Lambda = \Lambda (s^2,n_{pp}^2,n_{nn}^2,n_{sp}^2,n_{sn}^2,n_{pn}^2) \, ,
\end{equation}
so that the most general form of the momenta is
\begin{equation} 
\label{mometumungauged}
\begin{array}{l}
\mu^n_{\nu} = \mathcal{B}^n n_{n\nu} + \mathcal{A}^{sn} s_\nu  + \mathcal{A}^{pn} n_{p\nu}
\\
\mu^p_{\nu} = \mathcal{B}^p n_{p\nu} + \mathcal{A}^{sp} s_\nu  + \mathcal{A}^{pn} n_{n\nu}
\\
\Theta_{\nu} = \mathcal{C} s_{\nu} + \mathcal{A}^{sp} n_{p\nu}  + \mathcal{A}^{sn} n_{n\nu}  \, .
\end{array}
\end{equation}
In the absence of dissipation we must impose $s_\nu = s  \, n_{p\nu}/n_p$ and the above definitions become
\begin{equation} 
\label{mometumungauged2.0}
\begin{array}{l}
\mu^n_{\nu} = \mathcal{B}^n n_{n\nu} +\left( \mathcal{A}^{sn} \dfrac{s}{n_p}  + \mathcal{A}^{pn} \right) n_{p\nu}\\
\\
\mu^p_{\nu} = \left( \mathcal{B}^p + \mathcal{A}^{sp} \dfrac{s}{n_p} \right) n_{p\nu}  + \mathcal{A}^{pn} n_{n\nu}\\
\\
\Theta_{\nu} = \left( \mathcal{C}\dfrac{s}{n_p} + \mathcal{A}^{sp} \right) n_{p\nu}  + \mathcal{A}^{sn} \, n_{n\nu} \, .
\end{array}
\end{equation}
Therefore, at this level every conjugate momentum is expected to depend on all the currents of the theory. 
However, a convenient gauge fixing can be performed as follows: the s-locking property of the normal component implies  
\begin{equation}
n_{sp}^2 = \sqrt{s^2 \, n_p^2} 
\qquad
\qquad
n_{sn}^2 = \dfrac{n_{pn}^2 \sqrt{s^2}}{\sqrt{n_p^2}} \, , 
\end{equation}
so that it is possible to define an alternative master function $\Pi$ as
\begin{multline}
\Pi(s^2, n_{pp}^2, n_{nn}^2 , n_{pn}^2 ) 
:= \\
\Lambda \bigg( s^2, n_{pp}^2, n_{nn}^2 ,  \sqrt{s^2 \, n_{pp}^2} , \dfrac{n_{pn}^2 \sqrt{s^2}}{\sqrt{n_{pp}^2}}, n_{pn}^2 \bigg) \, .
\end{multline}
Under the hypothesis that the model is defined by the specification of $\Lambda$, the use of $\Pi$ is completely admissible because both master functions produce same Lagrangian density $\mathcal{L}$. Now, the entrainment matrix is not gauge independent; in particular, the diagonal coefficients $\tilde{\mathcal{B}}^x$ obtained as derivatives of $\Pi$, are linked to the ones obtained from $\Lambda$ via
\begin{equation} 
\label{coefficientigaugiati}
\begin{array}{l}

\tilde{\mathcal{B}}^n =
-2 \, \dfrac{\partial \Pi}{\partial n_{nn}^2} 
= \mathcal{B}^n \, .
\\
\tilde{\mathcal{B}}^p =
-2 \, \dfrac{\partial \Pi}{\partial n_{pp}^2} 
= \mathcal{B}^p + \mathcal{A}^{sp} \dfrac{s}{n_p} - \mathcal{A}^{sn} \dfrac{s \, n_{pn}^2} {n_p^3} 
\\
\tilde{\mathcal{C}}= 
-2 \, \dfrac{\partial \Pi}{\partial s^2} 
= \mathcal{C} + \mathcal{A}^{sp} \dfrac{n_p}{s} + \mathcal{A}^{sn} \dfrac{n_{pn}^2}{s \, n_p} 
\end{array}
\end{equation}
Regarding the off-diagonal coefficients, $\Pi$ does not depend on $n_{sp}^2$ and $n_{sn}^2$, implying that 
\begin{equation}
\tilde{\mathcal{A}}^{sp} = \tilde{\mathcal{A}}^{sn}=0 \,. 
\end{equation}
The entrainment is now described only in terms of a single anomaly coefficient:
\begin{equation}
\tilde{\mathcal{A}}^{pn}= -\dfrac{\partial \Pi}{\partial n_{pn}^2} = \mathcal{A}^{pn} + \mathcal{A}^{sn} \dfrac{s}{n_p}.
\end{equation}
The calculation of the new momenta is straightforward
%
%
and  it is possible to verify that  
\begin{equation} 
\label{mometigaugiati2.0}
\begin{array}{l}
\tilde{\mu}^n_{\nu} = \mu^n_{\nu}\\
\\
\tilde{\mu}^p_{\nu} = \mu^p_{\nu} +  \dfrac{s}{n_p} \mathcal{A}^{sn} \bigg[ n_{n\nu} - \dfrac{n_{pn}^2 n_{p\nu}}{n_p^2}  \bigg]\\
\\
\tilde{\Theta}_{\nu} = \Theta_\nu - \mathcal{A}^{sn} \bigg[ n_{n\nu} - \dfrac{n_{pn}^2 n_{p\nu}}{n_p^2}  \bigg] \, .
\end{array}
\end{equation}
The above system of equations is obviously consistent with the general form in \eqref{gaugeingenerale}, where
\begin{equation}
\pi_\nu = \dfrac{s}{n_p} \mathcal{A}^{sn} \bigg[ n_{n\nu} - \dfrac{n_{pn}^2 n_{p\nu}}{n_p^2}  \bigg]
\end{equation} 
and
\begin{equation}
\zeta_\nu = -\dfrac{n_p}{s} \pi_\nu = - \mathcal{A}^{sn} \bigg[ n_{n\nu} - \dfrac{n_{pn}^2 n_{p\nu}}{n_p^2}  \bigg].
\end{equation}
The result of the gauge transformation in \eqref{mometigaugiati2.0} can be written in a more elegant way with the aid of the two complementary projectors defined in \eqref{proettoriferoci1},
\begin{equation} 
\label{mometacomparison}
\begin{array}{l}
\tilde{\mu}^n_{\nu} = \mu^n_{\nu}\\
\\
\tilde{\mu}^p_{\nu} = \mu^p_{\nu} +  \dfrac{s}{n_p} ({\perp}\Theta)_\nu\\
\\
\tilde{\Theta}_{\nu} = (\paral \Theta)_\nu  \, .
\end{array}
\end{equation}
Thanks to this more geometrically transparent form, it is easy to notice that, since
\begin{equation}
\pi_\nu = \dfrac{s}{n_p} ({\perp}\Theta)_\nu \, ,
\end{equation}
also the first relation of \eqref{bazinga} is automatically satisfied, as it should be. Moreover, from the last expression in \eqref{mometacomparison}, it is also evident that we have redefined the entrainment to satisfy \eqref{carterizzoilthermal}. This completes the proof that, given an arbitrary $\Lambda$ and provided that $n_p \neq 0$, it is always possible to make a gauge-fixing which removes the entrainment from the entropy. 

Finally, it is possible to verify with direct calculations that $\Theta$, $\mu_x^{(T)}$, $Y_{nn}$, $\Psi$, $T^{\nu \rho}$ and the hydrodynamic equations are left unchanged by the above transformations. This proves that $\Lambda$ and $\Pi$ produce the same  hydrodynamic model when the LAB formalism is used.

The importance of this example is represented by the fact that it shows how the gauge can be used to simplify the hydrodynamic model, removing the redundancies. A further practical application that exploits the  gauge freedom of the system is presented at the end of section \ref{Chemical reactions in almost-perfect multifluids}.

\subsection{Constructing the master function from a given equation of state}
\label{costruireLambda}

We now present two simple examples of how to assign $\Lambda(n_{xy}^2)$ for a system described by a given equation of state. For simplicity, we assume to know $\mathcal{E}$ (if the microscopic analysis of the system provides a different thermodynamic potential it is sufficient to perform a Legendre transformation). 
\\
\\
\emph{Tisza-Landau two-fluid model} ($l=0$, $k=1$) 
\\
\cite{cool1995} proposed a relativistic version of the Tisza-Landau model for Helium-II: we briefly discuss this model for a single component superfluid at finite temperature and provide a practical formula to set the master function.  

The two currents of the theory are  $s^\mu$, describing the gas of excitations, and a free current $n^\mu$. In general,  the master function has the form
\begin{equation}
\Lambda = \Lambda (s^2,n^2,n_{sn}^2) \, ,
\end{equation}
so that there is no gauge freedom here: $n_{sn}^2$ carries all the information about the relative speed between the two components. All we need to do is to rearrange the variables in order to make everything dependent only on three scalars that could be conveniently used in a thermodynamic potential. 
%
Let $n^\mu = n^0 e^\mu_0+n^1 e^\mu_1$ be the particle current  expressed in the frame in which excitations have average zero velocity; the potential $\mathcal{E}$ may be given as a function of three thermodynamic variables as 
\begin{equation}
\label{eq:E_he4}
\mathcal{E} \, = \, \mathcal{E}(s^0,n^0,n^1) \, .
\end{equation} 
Noticing that 
\begin{equation}
s^0\,  = \, \sqrt{s^2}  
\qquad \qquad
n^0 \, = \, n_{sn}^2 \, / \, \sqrt{s^2} \,  ,
\end{equation} 
it is immediate to write $n^1$ in terms of the fundamental scalars $s^2$, $n^2$ and $n_{sn}^2$ as
\begin{equation}
n^1 = \sqrt{(n^0)^2 - n^2} \, .
\end{equation}
We conclude that, given the thermodynamic potential in \eqref{eq:E_he4}, a consistent hydrodynamic theory can be constructed starting from the master function
\begin{equation}
\Lambda (s^2,n^2,n_{sn}^2) 
\, = \,  
-\mathcal{E}\bigg(\sqrt{s^2},\dfrac{n_{sn}^2}{\sqrt{s^2}}, \sqrt{\dfrac{(n_{sn}^2)^2}{s^2} - n^2} \bigg) \, .
\end{equation}
Alternatively, if the internal energy is provided, say $\mathcal{U}= \mathcal{U}(s,n^0,\mu_1)$, the first step to obtain the master function is to perform the Legendre transform $\mathcal{E}=\mathcal{U}-n^1 \mu_1$ and then to make the change of variables presented above.
\\
\\
\emph{Andreev and Bashkin model} ($l=0$, $k=2$) 
\\
We now consider the mixture of two superfluid species at finite temperature. Such a system was early considered by \cite{khala57} as a multicomponent generalization of the Tisza-Landau model, refined by  \cite{andreevbashkin1976}
for a superfluid $^3$He-$^4$He  mixture and used to construct a Newtonian description of the matter in the outer core of a neutron star by \cite{mendell1991}. We label the two species by $p$ and $n$; the corresponding currents, say $n_p^\nu$ and $n_n^\nu$, are both free (again, there is no gauge freedom since $l=0$). In this case, the general form of a master function is
\begin{equation}
\Lambda 
\,  = \, 
\Lambda (  s^2, \, n_{pp}^2,  \, n_{nn}^2,  \, n_{sp}^2,  \, n_{sn}^2,  \,n_{pn}^2 )  \, .
\end{equation}
It is convenient to choose a tetrad comoving with the gas of excitations such that the number of non-zero components of the two free currents is minimized, say $n_p^2 = n_p^3 = n_n^3=0$. Hence, let us suppose that an equation of state of the form 
\begin{equation}
\mathcal{E}
\,  = \, 
\mathcal{E}( s^0, n_p^0, n_n^0, n_p^1, n_n^1, n_n^2) \, 
\end{equation}
is given as the result of a microscopic study of the system. Similarly to the previous example, we perform the change of variables
\begin{equation}
\label{tostizeri}
 s^0 \, = \, \sqrt{s^2} 
\qquad 
 n_p^0 \, = \, n_{sp}^2 / \sqrt{s^2}  
\qquad 
 n_n^0 \, = \, n_{sn}^2 / \sqrt{s^2}
\end{equation}
and
\begin{equation}\label{tostod1}
n_p^1 \, = \, \sqrt{\dfrac{(n_{sp}^2)^2}{s^2} - n_{pp}^2}.
\end{equation}
The computation of $n_n^1$ and $n_n^2$ requires some extra effort. Since the projector ${\perp}$ introduced in \eqref{proettoriferoci1} can be rewritten as
\begin{equation}
{\perp}\indices{^\nu_\rho} 
\, = \,  
\delta\indices{^\nu _\rho} + \dfrac{s^\nu s_\rho}{s^2} \, ,
\end{equation}
it is immediate to find
\begin{equation}
({\perp}n_n)^a = (0,n_n^1,n_n^2,0) \, .
\end{equation}
A contraction with $-n_{pa}$ gives
\begin{equation}
n_{pn}^2 -\dfrac{n_{sn}^2 n_{sp}^2}{s^2} \, = \, - n_n^1 n_p^1 \, ,
\end{equation}
which can be casted into the form
\begin{equation}\label{tostor1}
n_n^1 \, = \,  \dfrac{n_{sn}^2 n_{sp}^2 - s^2 n_{pn}^2}{\sqrt{s^2} \sqrt{n_{sp}^4 -n_{pp}^2 s^2}}.
\end{equation}
Finally, we use the identity
\begin{equation}
n_{nn}^2 = (n_n^0)^2 - (n_n^1)^2 - (n_n^2)^2 
\end{equation}
to obtain
\begin{equation}\label{tostor2}
n_n^2 = \sqrt{ \dfrac{n_{sn}^4}{s^2} - \dfrac{(n_{sn}^2 n_{sp}^2 - s^2 n_{pn}^2)^2}{s^2 (n_{sp}^4 -n_{pp}^2 s^2)} -n_{nn}^2    }.
\end{equation}
We conclude that the covariant hydrodynamic theory for a mixture of two superfluid species generaed by the equation of state $\mathcal{E}$ can be constructed starting from the master function
\begin{equation}
\Lambda (s^2,n_{pp}^2,n_{nn}^2,n_{sp}^2,n_{sn}^2,n_{pn}^2) 
= 
- \mathcal{E}(s^0,n_p^0,n_n^0, n_p^1,n_n^1,n_n^2) \, ,
\end{equation}
where the arguments of $\mathcal{E}$ are seen as functions of the scalars $n_{xy}^2$ by means of \eqref{tostizeri}, \eqref{tostod1}, \eqref{tostor1} and \eqref{tostor2}.

This hydrodynamic model is the basis  for many models of  neutron star interiors, see  e.g. \cite{Comer_Langlois1999}. Even if a proper description should be built starting from (at least)  a $(l=1,k=2)$ multifluid that includes the normal electrons, the superconducting protons and the superfluid neutrons, see e.g. \cite{Gusakov2016}, a two-fluid description is often preferred. For example, \cite{KP2017} consider a Newtonian effective $(l=0,k=2)$ uncharged fluid at zero temperature,  in which one current is comprised of superfluid neutrons and the other one is a superfluid  protons in a neutralizing bath of electrons that plays no dynamical role.
The same approach has been used also by \cite{pca2002}, where a useful Newtonian formalism to account for entrainment in superfluid neutron stars has been developed.
In appendix \ref{kasumotomorizuma} we show how to translate the Lagrangian formalism of Prix and collaborators into the Hamiltonian one developed by \cite{KP2017}, showing that this  operation is more transparent if one considers both the descriptions as low-energy limits of the relativistic multifluid theory.

\section{Chemical reactions}

The thermodynamic language developed in the previous sections is a useful tool not only for the description of perfect multifluids, but also to discuss some physical situations in which dissipation occurs. In section \ref{Chemical reactions in almost-perfect multifluids} we review the covariant treatment of  dissipation that occurs when matter is transformed from one constituent to another due to chemical disequilibrium developed by \cite{carter1991RSPSA} and applied to beta reactions in neutron star interiors in \cite{langlois98}, see also \cite{carterXmilneXIII} for an analogous formulation in a Milne-Cartan framework. 
%
%
In section \ref{bellofigo}, as a practical application, we invoke the gauge freedom to simplify the rocket term present in the hydrodynamical model of \cite{langlois98} and use the result to prove that in the low temperature limit it reduces to the one considered by \cite{ColucciRocket2013}.
Finally, in section \ref{Thermodynamic approach}, we try to rederive the Carter and collaborators's  approach from a purely thermodynamical argument, providing a justification and insight for this treatment of dissipation induced by chemical reactions. 

%
%

\subsection{Chemical reactions in almost-perfect multifluids}
\label{Chemical reactions in almost-perfect multifluids}

Consider the generic perfect multifluid presented in section \ref{Local thermodynamics of perfect multifluids} and imagine to switch on a chemical reaction, defined by \eqref{lareazioncionagenerica}. 
If the reaction is  slow compared to all the other equilibriation processes, such as the friction between the normal components, we can impose that the multifluid is still in local thermodynamic equilibrium with respect to every degree of freedom apart from the chemical fractions \citep{haensel_frozen}. Within this scenario, we can work under the  assumption that all the normal currents are still s-locked and that viscous effects are negligible.
We refer to this substance as an almost-perfect multifluid: the only form of dissipation is due to chemical reactions, there is no shear viscosity and no heat flow is measured in the frame of the normal components. In other words, the normal components still behave as a mixture that can be modelled as a perfect fluid but the presence of chemical reactions alters the chemical fractions (of both normal and superfluid species) and produces a gain in entropy.
In this physical situation, the energy-momentum tensor keeps the form \eqref{energiamomentotempfinita}, which implies the following relation
\begin{equation}
\nabla_\rho T\indices{^\rho _\nu} = \sum_x [\mu^x_\nu\nabla_\rho n_x^\rho + 2n_x^\rho \partial_{[\rho}\mu^x_{\nu]}].
\end{equation} 
Since the equations $\nabla_\rho T\indices{^\rho _\nu}=0$ and $\partial_{[\rho}\mu^i_{\nu]}=0$ are still valid and that $n_A^\rho = n_A u^\rho$, we obtain
\begin{equation}\label{dwodwoedmwemod}
\sum_x \mu^x_\nu\nabla_\rho n_x^\rho + u^\rho \sum_{A=0}^l 2n_A \partial_{[\rho}\mu^A_{\nu]} =0.
\end{equation}
A contraction of the above equation with $u^\nu$ gives
\begin{equation}\label{delmaialenonsibuttavianiente}
\Theta \nabla_\nu s^\nu = -\sum_{x \neq s} \mu_x^{(T)} \nabla_\nu n_x^\nu.
\end{equation}
In section \ref{chemicandodurissimooo} we introduced the reaction coordinate $N=n^0 V$ as a global quantity which counts the net number of reactions from the left to the right. Now, in a hydrodynamical framework, we need to introduce an associated local analogue. A convenient way to do this is to employ the method of the 
\emph{virtual charges}: 
consider \eqref{lareazioncionagenerica} and imagine to add a fictitious particle of charge $+1$ any time a reaction from the left to the right occurs (similarly, a particle of charge $-1$ is added when the reaction proceeds in the opposite direction). These fictitious particles are just tools to keep track of the reactions: they appear in the point of the spacetime where the reaction event happens and, once created,   are supposed to be advected by the entropy current. 
 
We define the \emph{reaction current} $n^\nu$ to be the particle number current of the $+1$ charges minus the one of the $-1$ charges. Since  $n^\nu = n u^\nu$, the net density of fictitious charges $n$ coincides with  the quantity $n^0$ that was introduced in \ref{chemicandodurissimooo}. This is a useful mathematical construction because, for any $x\neq s$, it allows us to define the conserved currents 
\begin{equation}
J_x^\nu := n_x^\nu - (\beta_x - \alpha_x)n^\nu .
\end{equation}    
The conservation laws  $\nabla_\nu J_x^\nu=0$ are just the local differential version of equation \eqref{conservotutto} and can be used to rewrite \eqref{delmaialenonsibuttavianiente} as
\begin{equation}\label{propropropro}
\Theta \nabla_\nu s^\nu =  \sum_{x \neq s} ( \alpha_x  - \beta_x) \mu_x^{(T)} \nabla_\nu n^\nu.
\end{equation}
The left-hand side describes the production of heat as a result of chemical reactions. 
The second principle of thermodynamics and the requirement of causality  impose that $\nabla_\nu s^\nu \geq 0$, which allows us to write
\begin{equation}\label{ewoiodwoifmeripopemifvdfvmrp}
\begin{split}
& if \quad \sum_{x \neq s} \alpha_x \mu_x^{(T)} > \sum_{x \neq s} \beta_x \mu_x^{(T)} \quad \Longrightarrow \quad \nabla_\nu n^\nu \geq 0, \\
& if \quad \sum_{x \neq s} \alpha_x \mu_x^{(T)} < \sum_{x \neq s} \beta_x \mu_x^{(T)} \quad \Longrightarrow \quad \nabla_\nu n^\nu \leq 0. \\
\end{split}
\end{equation}
This is just the statement that the reactions proceed from the side having higher chemical potential to the other one: \eqref{ewoiodwoifmeripopemifvdfvmrp} is the local, covariant version of Le Ch\^atelier's principle. 
Clearly, when chemical equilibrium is reached, see \eqref{tantovalagatta...amorire!}, the reactions in the two directions are balanced, implying that there is no net production of fictitious charge,
\begin{equation}
if \quad \sum_{x \neq s} \alpha_x \mu_x^{(T)} = \sum_{x \neq s} \beta_x \mu_x^{(T)} \quad \Longrightarrow \quad \nabla_\nu n^\nu = 0 \, .
\end{equation}
Following \cite{langlois98}, we assume that  not too far from chemical equilibrium  the reaction rates are proportional to the unbalance between the chemical potentials (to ensure the validity of Le Ch\^atelier's principle, the constant of proportionality  $\Xi$ must be a positive coefficient). Rewriting \eqref{dwodwoedmwemod} using the conservation of $J_x^\nu$ we obtain 
\begin{equation}\label{thesystemrocket}
\begin{split}
& \nabla_\nu n^\nu = \Xi \sum_{x \neq s} (\alpha_x - \beta_x) \mu_x^{(T)}  \, =: \, r\\
& \Theta \nabla_\nu s^\nu = \Xi \bigg[ \sum_{x \neq s} (\alpha_x - \beta_x ) \mu_x^{(T)}  \bigg]^2 \\
&  \sum_{A=0}^l 2n_A^\rho \partial_{[\rho}\mu^A_{\nu]} = - \Theta_\nu \nabla_\rho s^\rho + \sum_{x \neq s} (\alpha_x - \beta_x)\mu^x_\nu \nabla_\rho n^\rho. \\
\end{split}
\end{equation}
The right-hand side of the first equation is called \textit{reaction rate} and will be  denoted by $r$ in the following. The new quantity $r$ describes the coupling between the hydrodynamic and the chemical evolution of the multifluid. 


The presence of superfluid currents has an interesting  consequence on the evolution of the system: a relative motion between the species gives rise to  an extra term in the hydrodynamic equations, the so-called \emph{rocket effect}, see \cite{prix2004} and \cite{ColucciRocket2013}. In fact, the third equation of \eqref{thesystemrocket} can be seen as an equation of motion for the variable $u^\nu$: the vector
\begin{equation}
\label{rocketman}
\mathcal{R}_{\nu} :=  - \Theta_\nu \nabla_\rho s^\rho + r \sum_{x \neq s} (\alpha_x - \beta_x)\mu^x_\nu \, .
\end{equation}
describes the acceleration of the normal components. This rocket effect results from the fact that the 
transfusion process produces an overall redistribution of the momenta between the various species. 
Notice that $\mathcal{R}_{\nu}$ vanishes when all the species comove, as can be seen by using \eqref{propropropro} with the assumption that $\Theta_\nu = \Theta u_\nu$  and $\mu^x_\nu = \mu_x^{(T)}u_\nu$.

\subsection{How to model chemical reactions in the zero temperature limit}
\label{bellofigo}

Under a gauge transformation of the type \eqref{gaugeingenerale}, the  system of dynamical equations \eqref{thesystemrocket} written in therms of the gauge-fixed momenta must be equivalent to the original one. In fact, to derive it we only invoked the the energy-momentum conservation and the irrotationality of the superfluid momenta, which are gauge invariant properties of the system. 
However, the two sides of the third equation in \eqref{thesystemrocket} are not separately gauge invariant: the rocket term  transforms as
\begin{equation}
\tilde{\mathcal{R}}_\nu = \mathcal{R}_\nu - \sum_{A=1}^l r_A \pi^A_\nu,
\end{equation}
with
\begin{equation}
r_A := s^\rho \nabla_\rho \bigg( \dfrac{n_A}{s} \bigg) = (\beta_A -\alpha_A)r - \dfrac{n_A}{s \Theta} \dfrac{r^2}{\Xi}.
\end{equation}
Furthermore, it is possible to check that  the left-hand side of the third equation of \eqref{thesystemrocket} changes in the same way,
\begin{equation}
 \sum_{A=0}^l 2n_A^\rho \partial_{[\rho}\tilde{\mu}^A_{\nu]} =  \sum_{A=0}^l 2n_A^\rho \partial_{[\rho}\mu^A_{\nu]}  - \sum_{A=1}^l r_A \pi^A_\nu,
\end{equation}
ensuring that a gauge transformation leaves the overall equation unchanged, as  expected.

Realizing that the rocket term $\tilde{\mathcal{R}}_\nu$ is a gauge dependent quantity is particularly important in the modelling of chemical reactions in neutron star interiors in the zero temperature limit, as done e.g. by \cite{ColucciRocket2013}. Strictly speaking, chemical reactions cannot be consistently introduced in a zero temperature formalism simply because, as it is evident from the second equation in \eqref{thesystemrocket}, they are associated to an increase of entropy. However, it is possible to perform a gauge fixing  in a way that  \eqref{carterizzoilthermal} holds; in this way  equation \eqref{rocketman} becomes
\begin{equation}
\mathcal{R}_{\nu} = r \sum_{x \neq s} (\alpha_x - \beta_x) {\perp}\indices{^\sigma _\nu}\mu^x_\sigma ,
\label{rocketzero}
\end{equation}
where the dependence on $\nabla_\rho s^\rho$ has been eliminated. On the other hand, with this gauge choice we have that $\Theta_\nu = \Theta u_\nu$, which goes to zero in the low temperature limit\footnote{
With a different gauge fixing the thermal momentum $\Theta_\nu$ may not go to zero in the low temperature limit. Instead, it may diverge in order to keep the product $s\Theta_j$ finite, as can be  seen from equation \eqref{bazinga} for $\zeta_\nu$.
}. 
It follows that it is possible to drop the the $A=0$ term in the sum on the left hand side of  \eqref{thesystemrocket}: this provides a justification of the zero temperature form of the rocket effect used  in the analysis of neutron star oscillations made  by \cite{ColucciRocket2013}. 

Equation \eqref{rocketzero} is the relativistic version of rocket term derived by \cite{prix2004} in a Newtonian framework (see equation (177) therein)  under the assumption of zero entrainment with the entropy. 
%
However, we have proven that it is always possible to impose the condition \eqref{carterizzoilthermal} by means of a gauge fixing, so that there is no need to invoke it as a symplifying assumption.

\subsection{Thermodynamic approach   }
\label{Thermodynamic approach}

%

In this subsection we give an alternative derivation of the Le Ch\^atelier's principle \eqref{ewoiodwoifmeripopemifvdfvmrp} and of the first two equations in \eqref{thesystemrocket} following a purely thermodynamical approach. Our aim is to convince the reader that the common scheme adopted in chemistry to study reactions in a mixture of normal species \citep[see e.g. ][]{prigogine1969chemical} is not modified by the presence of currents associated to superfluid species.

Let us consider a box in contact with an ideal heat bath, as described in section \ref{EwHb}. 
Assume that the multifluid has been prepared in an homogeneous initial configuration and that the reaction is sufficiently slow compared  other equilibriation processes. More precisely, the reactions have to be so slow that  evolution of the system is a quasistatic transformation moving across the manifold $\gamma \subset \mathcal{Q}$ of the homogeneous macrostates in which  every degree of freedom, apart from the chemical fractions, assumes its equilibrium value. 
This construction formalizes the concept of quasi-equilibrium described in \cite{Yakovlev2001PhR}, see also  \cite{haensel_frozen}.

For simplicity, we consider a single reaction channel, defined by a particular relation of the type \eqref{lareazioncionagenerica} and involving an arbitrary number of components. 
Hence, $\gamma$ must be a one-dimensional manifold and can be parametrised by a single variable: for this purpose it is natural to choose $n^0$, as it is the parameter that measures the extent in which the reaction proceeds. 
In particular, remembering that $(n^0,s^0,\Theta_j,\mu_j^x)_{x \in \mathcal{N}}$ is the full set of the $5+3l$ independent variables of state that are compatible with the constants of motion, see section \ref{chemicandodurissimooo}, the curve $\gamma$ is  implicitly defined by the $4+3 l$ constraints
\begin{equation}
\label{einvewrpcompoe}
s^j =0  \spc n_A^j =0  \spc  \Theta = \Theta_H \, .
\end{equation}
In principle, these constraints can be used to  parametrise $\Theta_j$, $\mu_j^A$ and $s^0$ as functions of $n^0$.

Since the system evolves along $\gamma$, all we need to know is the set of values $\mathcal{F}(n^0)$ assumed by the free energy on this curve. Combining equations \eqref{equilibration}, \eqref{mnnbufti} and \eqref{einvewrpcompoe}  with the fact that $\mu_j^i$ are constants of motion, it is possible to show that the differential of the free energy  on $\gamma$ reads
\begin{equation}
d\mathcal{F} = \sum_{x \neq s} (\beta_x - \alpha_x)\mu_x^{(T)} d n^0 , \, 
\end{equation}
which leads us to introduce the \textit{affinity} of the reaction\footnote{
The affinity is usually defined by means of the Gibbs free energy because in chemistry reactions typically occur at constant pressure. However, in section \ref{Equilibrium properties of homogeneous systems} we assumed a definite control volume $V$, so we work with the Helmholtz free energy.
}
\begin{equation}\label{mummia}
\mathbb{A} :=-\dfrac{d \mathcal{F}}{d n^0} = \sum_{x \neq s} (\alpha_x - \beta_x )\mu_x^{(T)}.
\end{equation}
If the minimum of $\mathcal{F}$ does not fall on the border of the domain of $n^0$ (as discussed in section \ref{chemicandodurissimooo}), the equilibrium is defined by the minimum of $\mathcal{F}$ over $\gamma$, i.e. when
$\mathbb{A}=0$. That is obviously just another way to write equation \eqref{tantovalagatta...amorire!}. 
Reinterpreting \eqref{mummia} in the view of the LAB formalism introduced in section \ref{bsgformalism}, it is immediate to show that $\mathbb{A}=0$ gives exactly the chemical equilibrium condition used by \cite{Gusakov2007} for an electron, proton and neutron mixture, cf. equation (44) therein. 
 
Now, let us suppose that the initial fractions are slightly out of equilibrium, so that we can expand $\mathcal{F}$ around  the equilibrium value of the extent parameter  $n^0_{eq}$,
\begin{equation}
\mathcal{F}(n^0) \approx \mathcal{F}_{eq} + \dfrac{1}{2} \mathcal{F}_{eq}'' (n^0 - n_{eq}^0)^2, \spc \mathcal{F}_{eq}'' > 0.
\end{equation}
In this case the affinity can be approximated as
\begin{equation}\label{affinity}
\mathbb{A} = \mathcal{F}_{eq}'' (n_{eq}^0 - n^0),
\end{equation}
The second principle of thermodynamics forces $n^0$ to evolve towards the equilibrium value, namely
\begin{equation}\label{lechateliers}
\begin{split}
& \mathbb{A}>0 \quad \Longrightarrow \quad n_{eq}^0 > n^0 \quad \Longrightarrow \quad \dfrac{dn^0}{dt} >0 
\\
& \mathbb{A}<0 \quad \Longrightarrow \quad n_{eq}^0 < n^0 \quad \Longrightarrow \quad \dfrac{dn^0}{dt} <0 \, , 
\end{split}
\end{equation}
where the precise meaning of the parameter $t$ is provided by the construction in section \eqref{EwHb}. 
Hence, this result is the homogeneous (purely thermodynamic) version of equation \eqref{ewoiodwoifmeripopemifvdfvmrp}.

To advance in the analysis we need to include some kinetic considerations about the speed of the reaction. 
As we said, the evolution of the system is assumed to be a sequence of states which are in equilibrium with respect to all the variables (but with assigned fractions), so that $r$ can be regarded as a thermodynamic variable itself, and we can write\footnote{
Now that  we are dealing with the kinetics of the reaction channel, we need to require that the presence of the heat bath does not alter the speed of reaction. This means that it does not act as a catalyzer nor as a  inhibitor.
}
\begin{equation}\label{lereazionicipiaccioniassai}
\dfrac{dn^0}{dt} = r(\gamma,n^0).
\end{equation}  
The dependence on the particular curve $\gamma$, defined by the initial state in which the system is prepared, implicitly defines the dependence on the constants of motion that are conserved during the evolution towards equilibrium. Now, if we assume that we are near equilibrium, it is possible use the equation \eqref{affinity} to write $r$ as a function of $\mathbb{A}$. An expansion of $r$ near $\mathbb{A}=0$ leads to  the formula
\begin{equation}\label{frankenstein}
r \approx  \Xi \, \mathbb{A} \, ,  \spc  \Xi = \Xi(\gamma) >0.
\end{equation}
We have obtained the homogeneous version of the first equation in \eqref{thesystemrocket}. Notice that this formula 
is in complete accordance with equations (45) and (50) of \cite{Gusakov2007} and equation (44) of \cite{langlois98},   
providing a purely thermodynamic justification for both approaches (that may seem different only because these two  studies have been developed by using two quite different languages).

To complete the analysis of the hydrodynamical scheme  proposed by Carter and collaborators, we have to combine \eqref{mummia}, \eqref{lereazionicipiaccioniassai} and \eqref{frankenstein} to find that the  heat produced by the chemical reactions contributes to the variation of free energy according to the law 
\begin{equation}
\dfrac{d \mathcal{F}}{dt} = -\Xi \mathbb{A}^2,
\end{equation}
which is in accordance with the second equation of \eqref{thesystemrocket}. 

There is an interesting final remark to make: in the above analysis we have assumed that the interaction with the heat bath drives the system towards thermal equilibrium with respect to the variables $s^0$ and $\mu^A_j$ faster than with respect to the chemical fractions. This neutralizes the action of the rocket term because, as a result of an ideally instantaneous damping of normal flows in the frame of the bath, the normal species are not allowed to accelerate. For this reason, the third equation in \eqref{thesystemrocket} cannot be derived in a purely thermodynamic study: it can arise only in an hydrodynamical treatment. 
The present analysis of an homogeneous system is, therefore, only complementary to the one of Carter and collaborators. We can exploit the absence of the rocket term in the present treatment to solve exactly equation \eqref{lereazionicipiaccioniassai}, finding a formula for the time-scale of the reactions, namely
\begin{equation}
\mathbb{A}(t)=\mathbb{A}(0) e^{-t/\tau_{eq}},
\end{equation}
where the equilibration time is
\begin{equation}
\tau_{eq} = \dfrac{1}{\Xi \mathcal{F}_{eq}''}.
\end{equation}
Therefore, the presence of the rocket term in the hydrodynamic formulation tells us that the dynamics of the reaction and the motion of the components are coupled on the same time-scale $\tau_{eq}$. This leads to the practical complication that these two effects cannot be studied separately unless all the species comove.
%

\section{Conclusions}
\label{sec:conclusions}

We have developed a thermodynamic language for a  mixture of superfluid and normal species 
in General Relativity, without any approximation regarding the smallness of the relative currents between the components.
Our strategy and findings can be summarized as follows:


i) The central idea of our analysis has been introduced in section \ref{EwHb}, where  we consider an homogeneous subsytem  in contact with an heat bath. We define  the total free-energy in terms of three parameters: the temperature of the bath and the entropy and energy of the subsystem, see \eqref{dacapoancoraallinizio}. 
The link with the phenomenological model of section \ref{Gftva} is in the association of the entropy with the density relative to  one of the currents  (that we called $s^\mu$ beforehand) and of the  energy with the appropriate component of the energy-momentum tensor, namely equations \eqref{belloo} and \eqref{bello}.  

ii) Additional links between the state variables of the homogeneous subsystem and the phenomenological model can be provided by considering the quantities conserved during the equilibration process.
Of particular interest is the constraint arising from the irrotationality requirement of the superfluid momenta: although the presence of persistent  currents breaks the concept of comoving reference frame, we find no difficulties in defining the thermodynamic properties of the system. 

iii) The absolute temperature  and the chemical potentials  are defined by considering the relevant physical process (the equilibration with a heat bath in section \ref{Constraining the variations in the non-transfusive limit} and the relaxation to chemical equilibrium in \ref{Constraining the variations in the non-transfusive limit}).

iv) The thermodynamic language developed for the homogeneous system is embedded into the non-homogeneous phenomenological model thanks to the tetrad formalism, leading to the LAB formalism developed by \cite{Son2001} and \cite{GusakovAndersson2006}.

vi) Finally, we described how the phenomenological approach acquires a gauge freedom when its dynamics is  resticted to field configurations that are consistent with local equilibrium macrostates. However, the thermodynamic quantities (i.e. the quantities that appear in the LAB formalism) turn out to be gauge invariant. This has interesting consequences on the entrainment matrix and we provided a concrete example of gauge fixing in \ref{A concrete example} by considering the neutron star crust model of \cite{langlois98}.
\\
\\
During our discussion we rediscovered some results already known in the literature, providing  further justification  that is complementary to the one in the original works. In particular, we derived a chemical equilibrium condition that is consistent with both the one proposed by \cite{langlois98} and the one of \cite{Gusakov2007}.
Furthermore, we have shown in section \ref{Thermodynamic approach} that the usual concepts of affinity and reaction coordinates, as they are introduced in chemistry, are not affected by the presence of superfluid currents.

Building on the present formalism, it is possible to  extend  our work in two main directions. 

First, following \cite{mendell1991} and \cite{carterKalb} it may be interesting to include the effect of quantized vortices providing a  description of the kind required for application to relativistic rotating neutron stars. 
The presence of a vorticity field can have interesting consequencies from the thermodynamic point of view, as it is a further degree of freedom that breaks the isotropy of the system and contributes to the energy density of the system.
In fact, the procedure to extract the statistical meaning of hydrodynamic quantities defines a rigorous way to study any additional degree of freedom in the equation of state. As we have shown here, in fact, it represents a simple strategy to provide an unambiguous microscopic interpretation to quantities appearing in phenomenological models revealing possible redundant information they may carry.

Secondly, it is tempting to extend the analysis carried out in section \ref{Chemical reactions in almost-perfect multifluids} to study the emergence of reaction-induced viscosity in relativistic models. In upcoming works, the same approach will be applied to the study of dissipation, shown to arise directly from the implementation of out-of-equilibrium degrees of freedom in the equation of state. If the ideas presented in \eqref{Thermodynamic approach} are properly extended, our framework has the potentiality to be applied to a wide range of problems in relativistic astrophysics, ranging from various explosive scenarios to the internal layers of neutron stars.

\section*{Acknowledgements}

We acknowledge support from the Polish National Science Centre grant SONATA BIS 2015/18/E/ST9/00577, P.I.: B. Haskell and from the PHAROS COST Action (CA16214). The authors thank B. Haskell and N. Andersson for reading the manuscript and critical comments. MA thanks the Institute for Nuclear Theory at the University of Washington for its kind hospitality. This research was supported in part by the INT's U.S. Department of Energy grant No. DE-FG02-00ER41132.


\appendix

\section{Extracting the master function from a RMF model}
\label{CconRMF}

In section \ref{costruireLambda} we considered two basic examples of how to obtain the master function from a given thermodynamic potential. Here we expand the discussion and sketch a strategy that can be used when the microscopic theory provides directly the energy-momentum tensor. This is the case of the relativistic mean field (RMF) technique, a method used to extract the equation of state starting from the  Lagrangian of the quantum fields of interest \citep{glendenning_book}. In this sense, the RMF is a  way to deal with the problem of how to connect the microscopic scale, where one speaks in terms of interacting boson and fermion fields, and the fluid scale, where matter is contiuous and described in terms of thermodynamic variables.

The RMF can be used to construct the internal energy of the system when there are no relative currents (this case $\mathcal{U}=-\Lambda=T^{00}$ in the rest frame of matter). If a single relative current  between the species is present into the system, obtaining $\Lambda$ is less trivial but still possible: indeed, \cite{comerRMF} generalized the RMF procedure for a mixture of superfluid neutrons and supercondicting protons in the outer core of a neutron star starting from a mean field model for the nucleons and their interactions.

In brief, in the RMF  the energy-momentum tensor $T^{\nu \rho}$ and the currents $n_x^\nu $ of the fluid are calculated as mean field averages of the relative microscopic counterparts. Then, $T^{\nu \rho}$ and $n^\nu_x$ have to be combined algebraically in order to extract $\Lambda(n_{xy}^2)$ and the thermodynamic variables of interest.
Here, we address this algebraic problem when there are three independent current fields $n_1^\nu$, $n_2^\nu$, $n_3^\nu$, namely two relative currents between the species. This situation can be obtained in three cases: $(l=0,k=3)$ in the zero temperature limit, $(l=0,k=2)$ at finite temperature and $(l=1,k=2)$ in the zero temperature limit. In the following, the species $2$ and $3$ are always considered superfluid; the three possibilities are distinguished by the nature of the species $1$ which may be a further superfluid, the entropy, or a normal component.
Suppose that the current 1-forms
\begin{equation}
n_x \, := \, n_{x\nu} \, dx^\nu  
\qquad for  \qquad
x = 1,2,3 \,.
\end{equation}
are linearly independent and define the 1-form $N$ as\footnote{
The symbol $*$ is the Hodge duality operator: for a generic 3-form $\omega$, the duality is defined as
$(*\omega)_\mu = \varepsilon\indices{^{\nu\rho\sigma}_{\mu}} \,  \omega_{\nu\rho\sigma} /3!$ and $\varepsilon\indices{_{\alpha\beta\gamma\delta}} = \sqrt{-g} \, \text{perm}(\alpha\beta\gamma\delta)$ is the volume 4-form. 
Since $(n_1 \wedge n_2 \wedge n_3)_{\nu\rho\sigma} \, = \, 3! \, n^1_{[\nu}n^2_{\rho}n^3_{\sigma]}$, we have that $N_\mu = \varepsilon\indices{_{\nu\rho\sigma\mu}} n_1^\nu n_2^\rho n_3^\sigma$.
The linear independence of the 1-forms $n_x$ ensures that $N$ is not zero.
}
\begin{equation}
N \, := \, *(n_1 \wedge n_2 \wedge n_3) \,  
\end{equation}
Using the properties of Hodge duality, it can be proven that
\begin{equation}
N_\nu \, n_x^\nu  \, =\, 0 
\qquad \qquad
N_\nu \, N^\nu \,=\,  \det[n_{xy}^2] \,,
\end{equation}
where $ \det[n_{xy}^2]$  is the determinant of the $3 \times 3$ symmetric matrix $n_{xy}^2$. 
Since $N$ is orthogonal by construction  to every current, we use it to extract the generalised pressure from the energy-momentum tensor:
\begin{equation}
T^{\nu \rho} N_\nu N_\rho  \,=\, \Psi \, \det[n_{xy}^2] \, .
\end{equation}
Taking the trace of the energy-momentum tensor,
\begin{equation}
T\indices{^\nu _\nu} \, = \, 3 \, \Psi + \Lambda \, ,
\end{equation}
we obtain that the master function can be constructed as
\begin{equation}
\Lambda(n_{xy}^2) 
\, = \, 
T\indices{^\nu _\nu} - 3 \, T^{\nu \rho} \, \dfrac{N_\nu \,  N_\rho}{ \det[n_{xy}^2]} \, .
\end{equation}
%
Thanks to this  method we can easily see what happens when the independent currents are more than three (i.e. the system can support more than two relative currents between the components): in this case there are are infinite possible algebraic ways of combining the currents to give the same energy-momentum tensor and it is not possible to use them to uniquely reconstruct $\Lambda$.
The reason is that with four linearly independent currents there is no field $N^\nu$ which is orthogonal to every $n_x^\nu$, implying that it is impossible to extract algebrically the generalized pressure by using only the energy-momentum tensor and the currents.

%

\section{Chemical gauge in an arbitrary perfect multifluid}
\label{Chemical gauge in an arbitrary perfect multifluid}

In this appendix we generalize the idea of chemical gauge introduced in \cite{carter_macro_2006} to a generic perfect multifluid. According to the Carter formalism, the fundamental fields used to formulate the hydrodynamic description are the currents $n_x^\nu$. As anticipated in section \ref{sec:Gftva}, a chemical species is a  subset (labelled by $x$) of the total amount of particles that, in genereal,  may not reflect the real chemical composition of matter  as was earlier pointed out by \cite{CarterKhalat92}. 
In principle, it is possible to redefine the fundamental currents of the theory using a different prescription to assign each particle to a certain subset $x$ (the entropy has not to be considered a chemical species, so it cannot be redistributed). 

Immagine now to start with a theory in which the particles have been grouped into $l+k$ subsets. The theory has a  chemical gauge freedom if it is possible to redistribute the particles (according to a fixed rule) among the subsets. 
Clearly, not every redistribution rule is allowable. However, let us start by considering the most general rule,
\begin{equation}
\label{chemgaugissimo}
\tilde{n}_{y}^\nu = \sum_{x \neq s} Z\indices{^x _{y}} n_x^\nu \, ,
\end{equation} 
where $Z$ is an invertible $(k+l)\times (k+l)$ matrix whose components are functions of all the scalars $n_{xy}^2$. The possibility to perform such a transformation simply reflects the freedom to choose the coordinates of the master function, so it does not alter the equations of motion. The matrix $Z\indices{^x _{y}}$, however, is not completely arbitrary because of the further (physically motivated) subdivision of the currents into  s-locked and   free: the redefined currents $\tilde{n}_A^\nu$ still have to behave like normal currents, while the new momenta $\tilde{\mu}_\nu^i$ must be related to their order parameters. 
To implement the fact that the subdivision into s-locked and free currents should be preserved, we have to enforce that
\begin{equation}
\tilde{n}_A^\nu 
= 
\tilde{n}_A u^\nu  \spc \dfrac{\partial \Lambda}{\partial \tilde{n}_i^\nu} 
= 
\dfrac{\partial \Lambda}{\partial n_i^\nu}.
\end{equation}
The first condition implies $Z\indices{^i _A}=0$. The second one is more subtle and explicitly reads
\begin{equation}
\sum_{x \neq s} (Z^{-1})\indices{^i _x} \mu^x_\nu + \sum_{x,y \neq s} \mu^x_\rho \tilde{n}^\rho_y \dfrac{\partial (Z^{-1})\indices{^y_x}}{\partial \tilde{n}_i^\nu} = \mu_\nu^i \, .
\end{equation}
The simplest way to satisfy the above equation is to require  $Z\indices{^i _h}= \delta\indices{^i _h}$ and that the full matrix $Z$ does not depend on the superfluid currents. Hence, equation \eqref{chemgaugissimo} reduces to
\begin{equation}
\begin{split}
& \tilde{n}_A^\nu = \sum_{B=1}^l Z\indices{^B _A} n_B^\nu \\
& \tilde{n}_i^\nu = \sum_{B=1}^l Z\indices{^B _i} n_B^\nu + n_i^\nu \\
\end{split}
\label{wert}
\end{equation}
with $Z\indices{^A _x}= Z\indices{^A _x}(n_B,s)$. There is a further constraint which is worth noticing, related to the conservation of the currents. Suppose that there is a conserved current $J^\nu$ and that each particle of the species $x$ carries a formal charge $q^x$. This gives the constraint
\begin{equation}
J^\nu = \sum_{x \neq s} q^x n_x^\nu = \sum_{{x} \neq s} q^{x} \tilde{n}_{x}^\nu \, .
\label{qwerty}
\end{equation}
A comparison with \eqref{chemgaugissimo} tells us that $\sum_{y}  Z\indices{^x _{y}}q^{y} = q^x$. However, considering that $Z\indices{^i_{x}}= \delta\indices{^i _{x}}$, we see that   the only non-trivial condition implied by \eqref{qwerty} is
\begin{equation}
\sum_{x \neq s} Z\indices{^A _x}q^x = q^A \, .
\end{equation}
Furthermore, to have complete chemical gauge invariance $Z\indices{^A_x}$ should be constant along the worldlines: only in this way $\nabla_\nu \tilde{n}_x^\nu$ vanishes, giving rise to a Lagrangian description which is completely equivalent to the one obtained using the $n_x^\nu$.  In particular, the enrgy-momentum tensor, as defined by \eqref{energiamomentotempfinita} is invariant with respect to such a change of chemical basis. 

To provide an explicit example, we show how  the transformation \eqref{wert} comes down to the chemical gauge transformation considered in \cite{carter_macro_2006}, see equation (2.23) therein. Carter and collaborators considered three species in the crust of a neutron star, two s-locked (the ions that constitute the crustal lattice, $I$, and the protons, $p$) and one free (the neutrons, $n$) and perform the change of variables, 
\begin{equation}
\begin{split}
& \tilde{n}_I^\nu = n_I^\nu \\
& \tilde{n}_p^\nu = a_c n_p^\nu \\
& \tilde{n}_n^\nu = (1-a_c)n_p^\nu + n_n^\nu, \\
\end{split}
\end{equation}
where $a_c=a_c(n_I,n_p)$ and the conserved current  $J^\nu$ is the total baryon current $n_b^\nu=n_p^\nu+n_n^\nu$ (namely $q^p=q^n =1, q^I=0$). Clearly, this transformation has the general form \eqref{wert}, where $Z$ must fulfill also the additional constraint \eqref{qwerty}.

\section{An example of Hamiltonian formulation in the Newtonian limit}
\label{kasumotomorizuma}

In this appendix we address the practical problem of how to translate two different Newtonian formulations, the one of \cite{pca2002} and the one of \cite{KP2017}, for the hydrodynamics of the outer core of a neutron star under conditions when both neutrons ($n$) and protons ($p$) are superfluid. Since the model has been formulated for matter at zero temperature, there is no entropy current, i.e. both formulations are consider two superfluid species at zero temperature  (i.e. $k=2$, $l=0$).
Our aim is to provide a clear example of how a fully covariant theory may be used to build a bridge between two Newtonian theories that are apparently different. Moreover, the construction of a clear dictionary between these two different approaches has its practical usefulness for the reader interested in the modelling of neutron star interiors.
%
%

\subsection{The Newtonian limit of the Carter formalism}

Let us briefly review, following \cite{prix2004}, how to obtain the Newtonian limit of the Carter 
formalism\footnote{
Strictly speaking the formalism of \cite{prix2004}, which generalizes the model of \cite{pca2002}, is the Newtonian counterpart of the relativistic hydrodynamical description of superfluid neutron stars of \cite{prix2005}. The difference with respect to the language adopted by Carter is mainly in the definition of the entrainment parameters:  the approaches of Carter and Prix are completely equivalent and it is possible to translate one into the other.
}. 
To make the discussion more concrete, we stick to the case ($k=2$, $l=0$) for two superfluid species at zero temperature. In fact, the zero temperature case with two independent currents is of practical interest as it provides  the basic framework to model neutron star interiors \citep[see e.g. ][]{andersson_comer2000, and_com_lang_2002, prix2005, sourie2016PhRvD, antonelli+2018}.
We start by writing the potential $\mathcal{E}$ as a function of the rest frame densities of protons $n_p$ and neutrons $n_n$  and of the relative speed $\Delta$, 
\begin{equation}
d\mathcal{E} = \mu_p^{(C)} dn_p + \mu_n^{(C)} dn_n + \alpha d\Delta^2, 
\end{equation}
where $\mu_x^{(C)}$ are the comoving chemical potentials  defined in \ref{comcomcomcomcocm}. It is easy to show that the quantity $\alpha$ is related to the anomaly coefficient via
\begin{equation}
\label{tritone}
\alpha 
\,=\, 
\dfrac{1}{2} \, \mathcal{A} \, n_{np}^2 \, \Gamma^2 
\,=\, 
\dfrac{1}{2} \, \mathcal{A} \, n_n \, n_p \, \Gamma^4 
\, ,
\end{equation}
where $\Gamma=n_{np}^2=(1-\Delta^2)^{-1/2}$ is the Lorentz factor associated to the relative motion between the two species. With the perspective of taking the Newtonian limit we write $\mathcal{E}$ as
\begin{equation}
\mathcal{E} = m  n_p +m  n_n + \tilde{\mathcal{E}} (n_p,n_n,\Delta^2) \, ,
\end{equation}
where $m$ is the mass per particle of the mixture (this definition of $m$ allows us to forget about the electrons that, in principle, should be present into the system and define a third fluid).
 The differential of $\tilde{\mathcal{E}}$ is, then,
\begin{equation}
\label{iltildo}
d \tilde{\mathcal{E}} \, = \, \mu_p^{ACP} dn_p + \mu_n^{ACP}dn_n + \alpha \, d\Delta^2 \, ,
\end{equation}
where
\begin{equation}
\label{comovingchemicalsnewtonian}
\mu_x^{ACP} \, = \,  \mu_{x}^{(C)} - m 
\end{equation}
is the non-relativistic version of the comoving chemical potential. 

Following the notation used by 
\cite{KP2017}, we denote with $\mu_x^{ACP}$ the chemical potential used in \cite{acp2004}, as well as in \cite{prix2004}. Considering that
\begin{equation}
n_x = \sqrt{(n_x^0)^2 - \vect{n}_x^2} \approx n_x^0 - \dfrac{1}{2} \dfrac{\vect{n}_x^2}{n_x^0} \, ,
\end{equation}
the total master function reads
\begin{equation}
\Lambda \approx  -m n_p^0- m n_n^0 + \dfrac{m  \vect{n}_p^2}{2 n_p^0} + \dfrac{m  \vect{n}_n^2}{2 n_n^0} 
- 
\tilde{\mathcal{E}} (n_p^0,n_n^0,\Delta^2) \, . 
\end{equation}
The Newtonian master function is defined as the relativistic one deprived of the rest-mass energy density contribution, namely
\begin{equation}
\label{rinormalizzoduro}
\tilde{\Lambda}:= \Lambda+ m n_p^0 +m n_n^0
\, = \, 
\dfrac{m \vect{n}_p^2}{2 n_p^0} + \dfrac{m \vect{n}_n^2}{2 n_n^0} - \tilde{\mathcal{E}} \, .
\end{equation}
Now, following definition \eqref{noivinciamosempre},
\begin{equation}\label{negan}
\begin{split}
& \mu^p_j = \dfrac{\partial \Lambda}{\partial n_p^j} = \dfrac{\partial \tilde{\Lambda}}{\partial n_p^j} = m  v_{pj} - \dfrac{2\alpha}{n_p^0} (v_{pj}-v_{nj})  \\
& \mu^n_j = \dfrac{\partial \Lambda}{\partial n_n^j} = \dfrac{\partial \tilde{\Lambda}}{\partial n_n^j} = m  v_{nj} - \dfrac{2\alpha}{n_n^0} (v_{nj}-v_{pj}),  \\
\end{split}
\end{equation}
where
\begin{equation}
v_x^j = \dfrac{n_x^j}{n_x^0}
\end{equation}
is the velocity of the species $x$. For later convenience we introduce also the symbol 
\begin{equation}\label{epsilonnn}
\epsilon_x := \dfrac{2 \alpha}{m  n_x} 
\end{equation} 
which allows us to rewrite \eqref{negan} in the form
\begin{equation}\label{grimes}
\mu^x_j = m [(1-\epsilon_x)v_{xj} + \epsilon_x v_{yj}] \spc x \neq y \, .
\end{equation}
The entrainment parameters $\alpha$ and $\epsilon_x$ have, in general, an explicit dependence on the relative speed $\Delta$. However, in  \cite{KP2017} the entrainment coefficients are supposed, coherently with a low velocity limit, to be functions only of the densities, so that we should perform a first order expansion around $\Delta^2=0$, namely (c.f. \ref{iltildo})
\begin{equation}
\label{yogi}
\tilde{\mathcal{E}} (n_p^0,n_n^0,\Delta^2) 
\, \approx \, \tilde{\mathcal{U}}(n_p^0,n_n^0) + \alpha(n_p^0,n_n^0) \, \Delta^2 \, ,
\end{equation}  
where the quantity $\tilde{\mathcal{U}}$ is the non-relativistic energy density for comoving species as measured in their common frame\footnote{ We recall that if all the species comove $\mathcal{E}$ coincides with $\mathcal{U}$, see \eqref{EEEEEEEEEEEEEEEEEEEEEEEEEEE}, so $\tilde{\mathcal{E}} (n_p^0,n_n^0,\Delta^2=0) = \tilde{\mathcal{U}}(n_p^0,n_n^0)$.}.

\subsection{Introducing the Hamiltonian}

As a first step, we introduce the concept of Hamiltonian for a two-fluid system (the multifluid generalization should be obvious). This is done by considering the energy-momentum tensor in a certain frame but, since in the model of Kobiakov and Pethick there is no entropy current, we are forced to use an  observer   $\mathcal{O}$ that is completely generic. The energy density measured in the frame  $\mathcal{O}$ can be obtained as
\begin{equation}\label{eq:hamiltonianenergy2.0}
\mathcal{H}_\mathcal{O} = T^{00} = -\Lambda + n_p^j \mu^p_j + n_n^j \mu^n_j ,
\end{equation} 
which  is the Legendre transform of $\Lambda$ with respect to $n_x^j$. Its differential reads 
\begin{equation}
d\mathcal{H}_\mathcal{O} = \mu^{p0} dn_p^0 + \mu^{n0} dn_n^0 + n_p^j d\mu^p_j + n_n^j d\mu^n_j .
\end{equation}
Again, before making the Newtonian limit we subtract the rest mass, defining the Hamiltonian density 
\begin{equation}
\label{rinormalizzomorbido}
\tilde{\mathcal{H}}_\mathcal{O} := \mathcal{H}_\mathcal{O} - m n_p^0 - m n_n^0 \, .
\end{equation}
Taking the differential of both sides it is evident that
\begin{equation}
\label{firstlawgenericrenormalised}
d\tilde{\mathcal{H}}_\mathcal{O} 
= 
\mu^{\mathcal{O}}_p dn_p^0 + \mu^{\mathcal{O}}_n dn_n^0 + n_p^j d\mu^p_j + n_n^j d\mu^n_j,
\end{equation}
where
\begin{equation}
\label{ubuntu}
\tilde{\mu}_{p}^\mathcal{O} = \mu^{p0} - m  \spc  \tilde{\mu}_{n}^\mathcal{O} = \mu^{n0}- m,  
\end{equation}
are the non-relativistic chemical potentials in the frame of $\mathcal{O}$, not to be confused with the thermodynamic \eqref{chemichiamolachimica} or the comoving \eqref{comcomcomcomcocm} ones. Finally,
\begin{equation}
\label{eq:hamiltonianenergyrenormalised}
\tilde{\mathcal{H}}_\mathcal{O}= -\tilde{\Lambda} + n_p^j \mu_{j}^p + n_n^j \mu_{j}^n \, .
\end{equation}
is the analogous of \eqref{eq:hamiltonianenergy2.0} in the Newtonian limit.

\subsection{Connection between the Lagrangian and Hamiltonian approaches}

Using the definition of the momenta in  \eqref{negan},it is possible to prove the  useful identity,
\begin{equation}
\label{useful} 
 \dfrac{1}{2} n_p^j \mu_{j}^p + \dfrac{1}{2} n_n^j \mu_{j}^n = \dfrac{m  \vect{n}_p^2}{2 n_p^0} + \dfrac{m  \vect{n}_n^2}{2 n_n^0} - \alpha \Delta^2  \, ,
\end{equation}
that, used with   \eqref{rinormalizzoduro} and \eqref{yogi}, allow to cast the energy density given in \eqref{eq:hamiltonianenergyrenormalised}  into the form
\begin{equation}
\label{chespettacoloragazzi!}
\tilde{\mathcal{H}}_\mathcal{O}
\, = \, 
\tilde{\mathcal{U}} + \dfrac{1}{2}  n_p^j \mu_{j}^p + \dfrac{1}{2} n_n^j \mu_{j}^n \, .
\end{equation}
This expression highligths the separation into a kinetic and an internal part, cf. with equation (10) in \cite{KP2017}. This separation lead naturally to define a new chemical potential for each species as
\begin{equation}
\mu_x^{KP} := \dfrac{\partial \tilde{\mathcal{U}}}{\partial n_x^0} 
\qquad
\text{for}
\quad
 x=n,p
\, ,
\end{equation}
which coincides with the thermodynamic chemical potential $\mu_x^{(T)}$ in the limiting case in which all species comove. The label $KP$ is used to stress that this is the notion of chemical potential used by \cite{KP2017}.

To complete the dictionary we need to relate the two formulations of entrainment: in \cite{prix2004} the notation \eqref{entroinentrainmentconte} is used, while  \eqref{invertito!!!} is preferred in \cite{KP2017}. Inverting \eqref{grimes}, it is immediate to obtain
\begin{equation}
n_x^j 
\, = \, 
\dfrac{n_{xx}^{KP}}{m} \mu^{xj} + \dfrac{n_{x y}^{KP}}{m} \mu^{yj} 
\qquad
\text{with}
\quad
 x,y =\{n,p\} 
\text{, } 
x \neq y
\, ,
\end{equation}
where the diagonal element of the $n_{x y}^{KP}$ entrainment matrix are
\begin{equation}
n_{pp}^{KP} = \dfrac{n_p^0 (1-\epsilon_n)}{1-\epsilon_p-\epsilon_n } \qquad n_{nn}^{KP} = \dfrac{n_n^0 (1-\epsilon_p)}{1-\epsilon_n-\epsilon_p } , 
\end{equation}
while the off-diagonal ones are
\begin{equation}
 n_{pn}^{KP} = -\dfrac{n_p^0 \epsilon_p}{1-\epsilon_p-\epsilon_n } \qquad 
 n_{np}^{KP} = -\dfrac{n_n^0 \epsilon_n}{1-\epsilon_p-\epsilon_n }.
\end{equation}
Here, the coefficients $n_{xy}^{KP}$ do not depend on the relative speed and obey, as it should be, to the symmetry condition $n_{np}^{KP}= n_{pn}^{KP}$. They, also satisfy, considering \eqref{epsilonnn}, the equations
\begin{equation*}
n_{pp}^{KP}+ n_{pn}^{KP} = n_p^0 \spc  n_{nn}^{KP}+ n_{np}^{KP} = n_n^0,
\end{equation*}
which can be derived directly from the Newtonian limit of the relativistic equation $T^{0j}=T^{j0}$, namely
\begin{equation}
n_p^0 \mu^p_j + n_n^0 \mu^n_j = m (n_{pj}+ n_{nj}).
\end{equation}
Now that we know how to translate the entrainment coefficients, we can come back to the Hamiltonian for the system and rewrite \eqref{chespettacoloragazzi!} in the form
\begin{equation}
\label{accazero44}
\tilde{\mathcal{H}}_\mathcal{O}
\, = \,  
\tilde{\mathcal{U}}(n_p^0,n_n^0) + \dfrac{n_p^0}{2m} \boldsymbol{\mu}_p^2 
+ \dfrac{n_n^0}{2m} \boldsymbol{\mu}_n^2 - \dfrac{n_{np}^{KP}}{2m} 
|\boldsymbol{\mu}_p - \boldsymbol{\mu}_n|^2 \, ,
\end{equation}
which, remembering equation \eqref{firstlawgenericrenormalised}, implies
\begin{equation}\label{xsdfgrt}
\tilde{\mu}_x^\mathcal{O} 
= 
\mu_x^{KP} + \dfrac{\boldsymbol{\mu}_x^2}{2m} - 
 \dfrac{\partial n_{np}^{KP}}{\partial n_x^0} 
\, \dfrac{|\boldsymbol{\mu}_p - \boldsymbol{\mu}_n|^2}{2m} \, .
\end{equation}
This relation allows to link the notion of chemical potential used by Kobyakov and Pethick to the chemical potentials in \eqref{ubuntu}.

\subsection{The problem of the chemical potentials}

In \cite{koby2018PhRvC}, the  Euler-like equations for the two-fluid system  derived in \cite{KP2017} are compared with those expected in the context of Prix's theory. Kobiakov finds out that they coincide if it is true that
\begin{equation}
\mu_x^{ACP} -\dfrac{1}{2} m \epsilon_x^2 \Delta^2 = \mu_x^{KP} -  \dfrac{\partial n_{np}^{KP}}{\partial n_x^0} \,
\dfrac{|\boldsymbol{\mu}_p - \boldsymbol{\mu}_n|^2 }{2m}
\end{equation}
This is a non-immediate translation problem whose solution can be conveniently found by considering that the two formulations are the Newtonian limit of the same relativistic theory.

Let us use equations \eqref{comovingchemicalsnewtonian} and \eqref{ubuntu} into the definition of the comoving chemical potentials \eqref{comcomcomcomcocm}:
\begin{equation}
\label{troppinomi}
m_x + \mu_x^{ACP} \, = \, (  m_x+ \tilde{\mu}_x^{\mathcal{O}}  )\,  u_x^0 - \mu_{xj} \, u_x^j \, .
\end{equation}
The Newtonian limit of the above equation is immediately obtained thanks to the fact
\begin{equation}
u_x^0 \approx 1 + \dfrac{1}{2} \vect{v}_x^2  \spc u_x^j \approx v_x^j,
\end{equation}
so that, neglecting higher orders and imposing the approximation $m_x \approx m$, equation \eqref{troppinomi} reduces to
\begin{equation}
\label{ululatoauuuu!}
\mu_x^{ACP} =  \tilde{\mu}_x^{\mathcal{O}} + \dfrac{1}{2} m \vect{v}_x^2 - \mu_{xj}v_x^j \, .
\end{equation}
Employing equations \eqref{xsdfgrt} and \eqref{grimes}, we finally arrive at
\begin{align}
\nonumber
\mu_x^{ACP} 
= & \,
\mu_x^{KP}- \dfrac{\partial n_{np}^{KP}}{\partial n_x^0} \,
\dfrac{|\boldsymbol{\mu}_p - \boldsymbol{\mu}_n|^2 }{2m} +
\dfrac{\boldsymbol{\mu}_x^2}{2m}  + \dfrac{m \vect{v}_x^2}{2}  - \mu_{xj}v_x^j 
\\ 
= &
\, \mu_x^{KP}-  \dfrac{\partial n_{np}^{KP}}{\partial n_x^0} \,
\dfrac{|\boldsymbol{\mu}_p - \boldsymbol{\mu}_n|^2 }{2m}
+\dfrac{1}{2} m \epsilon_x^2 \Delta^2 \, ,
\end{align}
which is what we wanted to prove.

\bsp
\label{lastpage}
\end{document}